\documentclass[11pt]{article}
 \usepackage{jheppub}

\usepackage{color}
\usepackage{amsmath}
\usepackage{verbatim}
\usepackage{subfigure}
\usepackage{acronym}

\usepackage{amsfonts}
\usepackage{amssymb}
\usepackage{mathrsfs}
\usepackage{graphicx}
\usepackage{multirow}
 \usepackage{slashed}
 \usepackage{epsfig,multicol,bbm}
 \usepackage{url}
 \usepackage{float}

\usepackage[font={small}]{caption}   

\definecolor{myred}{rgb}{0.7, 0, 0}
\definecolor{myblue}{rgb}{0, 0, 0.7}
\definecolor{mygreen}{rgb}{0.04, 0.7, 0.5}
\definecolor{mygray}{rgb}{0.1, 0.1, 0.1}

\hypersetup{colorlinks,citecolor=myred,linkcolor=myblue,urlcolor=myblue,linktocpage=true}

 \def\be   {\begin{equation}}   \def\ee   {\end{equation}}
 \def\ba   {\begin{array}}      \def\ea   {\end{array}}
 \def\bea  {\begin{eqnarray}}   \def\eea  {\end{eqnarray}}
 \def\bean {\begin{eqnarray*}}  \def\eean {\end{eqnarray*}}
 \def\nn{\nonumber}
 \def\bry{\begin{array}}
 \def\ery{\end{array}}

\numberwithin{equation}{section}

\begin{document}

\begin{flushright}
\footnotesize
DESY-22-190 \\
\end{flushright}
\color{black}

\title{
Dilaton at the LHC: Complementary Probe of Composite Higgs
}
\date{\today}

\author[a,b]{Sebastian Bruggisser,}

\affiliation[a]{Institut f\"{u}r Theoretische Physik, Universit\"{a}t Heidelberg, D-69120 Heidelberg, Germany}

\affiliation[b]{Department of Physics and Astronomy, Uppsala University, 75120 Uppsala, Sweden}

\emailAdd{bruggisser@thphys.uni-heidelberg.de}

\author[c]{Benedict von Harling,}

\affiliation[c]{Institut de F\'isica d'Altes Energies (IFAE), The Barcelona Institute of Science and Technology, Campus UAB, 08193 Bellaterra (Barcelona), Spain}

\emailAdd{benedictvh@gmail.com}

\author[d]{Oleksii Matsedonskyi,}

\affiliation[d]{DAMTP, University of Cambridge, Wilberforce Road, Cambridge, CB3 0WA, United Kingdom}

\emailAdd{alexey.mtsd@gmail.com}

\author[e,f]{G\'eraldine Servant}

\affiliation[e]{Deutsches Elektronen-Synchrotron DESY, Notkestr. 85, D-22607 Hamburg, Germany}
\affiliation[f]{II. Institute of Theoretical Physics, Universit\"{a}t Hamburg D-22761, Germany}

\emailAdd{geraldine.servant@desy.de}

\abstract{
The dilaton is predicted in various extensions of the standard model containing sectors with an approximate spontaneously-broken conformal invariance. As a Goldstone boson of a spontaneously broken symmetry, the dilaton can naturally be one of the lightest new physics particles, and therefore may be the first new physics imprint observed in collider experiments. 
In particular, it can arise in composite Higgs models which are often assumed to have approximate conformal invariance in the UV. The dilaton is then a composite state, generated by the same sector that produces the Higgs.
We continue the exploration 
of composite dilaton signatures 
at the LHC, using the latest experimental data and analysing the future detection prospects. We elaborate on the connection of the dilaton properties with the properties of the Higgs potential, clarifying in particular the relation between the scale relevant for electroweak fine tuning and the scale controlling the dilaton couplings. 
This relation is then used to derive the experimental sensitivity to the dilaton in natural composite Higgs scenarios, which reaches $\sim 3$~TeV in dilaton mass for generic parameter choices. At the same time, we show that dilaton searches are a complementary direction to probe Higgs boson compositeness, with the sensitivity comparable or exceeding that of Higgs coupling measurements.}

%
%

\maketitle


\section{Motivation}\label{sec:incon}

The dilaton is a well-motivated new scalar particle that could appear around the TeV scale. 
It can arise for example in composite Higgs models addressing the Higgs mass naturalness problem and their 5D holographic duals.\footnote{More precisely, the composite dilaton is dual to the {\it radion} of the Randall-Sundrum models \cite{Rattazzi:2000hs}.}
In this context, it can play a prominent role for dark matter phenomenology~\cite{Blum:2014jca,Kim:2016jbz,Baldes:2021aph}, serving as a portal to dark matter.
Another important property of the dilaton is that it naturally drives a strong cosmological first-order phase transition
\cite{Creminelli:2001th,Konstandin:2011dr,Azatov:2020nbe}, 
 leaving a very large signature in gravitational waves \cite{Randall:2006py}. 
 It is also a leading candidate for motivating supercooled phase transitions with unique cosmological implications such as cold baryogenesis \cite{Konstandin:2011ds}, baryogenesis from  strong CP-violation \cite{Servant:2014bla}, QCD-induced electroweak phase transition \cite{vonHarling:2017yew,Baratella:2018pxi}, intermediate low-scale inflationary stages \cite{Nardini:2007me,Konstandin:2011dr,Baratella:2018pxi} and modified dark matter abundances \cite{Baratella:2018pxi,Baldes:2021aph,Baldes:2020kam}.
Another intriguing feature of the dilaton is its possible link to flavour physics and the fermion mass hierarchy, and the possibility to induce  baryogenesis from varying Yukawa couplings~\cite{vonHarling:2016vhf,Bruggisser:2018mus,Bruggisser:2018mrt}.  The interplay of the dilaton potential with the Higgs potential 
and the consequences for the electroweak phase transition  and electroweak baryogenesis were studied in detail in \cite{Bruggisser:2018mus,Bruggisser:2018mrt,us:ewbg}.

The dilaton is  the Goldstone boson of spontaneously broken conformal invariance~\cite{cpr,Megias:2014iwa,Coradeschi:2013gda,Bellazzini:2013fga,Pomarol:2019aae}.
Featuring the Goldstone symmetry protection, the dilaton can be significantly lighter than other new physics states, with a mass suppressed by the parameters breaking the conformal invariance (e.g.~beta-functions of slowly running couplings), and hence could be the first directly observed new-physics particle. 
In particular, a light dilaton was obtained explicitly in certain 5-dimensional models which could be dual to 4-dimensional strongly-interacting theories~\cite{cpr,Megias:2014iwa,Coradeschi:2013gda,Bellazzini:2013fga,Pomarol:2019aae}; the appearance of a light composite resonance was also observed in the lattice simulation of QCD-like theories with a large number of fermionic flavours~\cite{LatticeStrongDynamics:2018hun,LatKMI:2016xxi}.
The collider phenomenology of the dilaton has attracted a lot of attention in the past~\cite{Csaki:2007ns,Foot:2007as,Goldberger:2007zk,Fan:2008jk,Vecchi:2010gj,Appelquist:2010gy,Hur:2011sv,Barger:2011nu,Barger:2011hu,Chacko:2012vm,Abe:2012eu,Coriano:2012nm,Bellazzini:2012vz,Cox:2013rva,Cao:2013cfa,Jung:2014zga,Blum:2014jca,Ahmed:2015uqt,Bellazzini:2015nxw,Bandyopadhyay:2016fad,Kim:2016jbz,Abu-Ajamieh:2017khi,Chakraborty:2017lxp,Ahmed:2019csf,Sachdeva:2019hvk}. In this work, we present an update of the experimental bounds on the dilaton by interpreting the latest LHC exclusion limits, and also show the HL-LHC sensitivity projection. 

Our analysis is focused on scenarios where the Higgs boson is a composite state and arises as the Goldstone boson of some spontaneously broken global symmetry~\cite{Panico:2015jxa}.  These composite Higgs models are motivated by the gauge hierarchy problem, and often assumed to feature an approximate conformal symmetry in the UV~\cite{Contino:2010rs}. This can result in a composite dilaton with ${\cal O}$(100 GeV) -- ${\cal O}$(1 TeV) mass, generated by the same sector which produces the Higgs. 
Since the main underlying motivation for a 
dilaton in this case is the electroweak (EW)  scale naturalness, it is suggestive to investigate in detail the implications of naturalness for the properties of the  
dilaton. We use a unified framework for the description of the composite dilaton and Higgs boson, based on the constraints imposed by the presence of spontaneously broken conformal and flavour symmetries of the new-physics sector. Additionally, we impose relations between different parameters derived from a large-$N$ expansion, assuming that the underlying new strongly-coupled theory is $SU(N)$ Yang-Mills. This minimal number of assumptions allows to draw valuable conclusions on the dilaton couplings and use them to confront this scenario with experimental data. 

The paper is organised as follows. We set up the effective description for the composite Higgs and dilaton in Section~\ref{sec:eft}. In Section~\ref{sec:colliderpheno} we list the  dilaton interactions which are most relevant for collider experiments, and use them to derive experimental bounds in Section~\ref{sec:colliderbounds}. We discuss our results in Section~\ref{sec:disc}.

\section{Effective Description for Dilaton+Higgs}\label{sec:eft}

\subsection{Dilaton Description}\label{sec:coupling}

We will assume that the composite sector is approximately scale-invariant in the UV, but contains operators whose coefficients slowly run with energy. 
Such a slow running eventually results in the confinement at the $\cal O$(TeV) scale, which can then be interpreted as a scale of spontaneous breaking of conformal invariance. 
The spontaneous breaking in turn results in 
an associated parametrically light Goldstone boson -- the dilaton~\cite{cpr,Coradeschi:2013gda,Bellazzini:2013fga}, whose VEV is related to the confinement scale.
The fact that the dilaton mass (suppressed by the size of the explicit scale-invariance breaking) can be parametrically lower than the mass of other composite new physics states, allows us to concentrate, for the purpose of dilaton collider phenomenology, on the dilaton interactions with only the standard model (SM) states. These will dictate both its production and decay. 
We will therefore consider an effective field theory (EFT) where the heavy composite states are integrated out. 

The dilaton couplings can be deduced from the energy scaling properties of different operators. Simplistically, the construction of the EFT Lagrangian can be understood as follows (see e.g.~\cite{Chacko:2012sy} and references therein for similar discussions on the dilaton EFT). 
Scale invariance is spontaneously broken if the theory features a scalar operator which gets a non-zero vacuum expectation value (VEV). In the limit of weakly broken conformal invariance the field excitations around this VEV should correspond to a light state -- the dilaton.
More precisely, we will use the parametrisation where the VEV of the dilaton field $\chi$ sets the scale of conformal-invariance breaking, 
\be
\langle \chi \rangle \equiv \chi_0 \ne 0,
\ee 
while the physical dilaton quanta correspond to the excitations above this background, $\delta \chi = \chi-\chi_0$.  Other states of the conformal sector acquire a mass $\propto \chi_0$, of order a few TeV, and will be integrated out from our EFT.
The EFT then contains the SM states (which we assume to not be part of the conformal sector) and the conformal sector states whose mass is protected by some symmetries, such as a dilaton, and the Higgs which we discuss in the next section. At the same time this EFT does not feature conformal invariance anymore. However, treating $\chi$ as a dilatation symmetry-breaking spurion, transforming as $\chi \to \kappa^{-1} \chi$ under dilatations $x_\mu \to \kappa \, x_\mu$,
we can derive the form of the EFT interactions, by requiring the operators to be formally scale-invariant. 
For example, assuming that the SM fermions acquire mass from the dilaton VEV we can derive their interactions with the dilaton
\be
{\cal L}_{m_\psi} =  m_\psi \bar \psi \psi \; \; \to \; \; m_\psi \frac {\chi}{\chi_0} \bar \psi \psi =  m_\psi \left\{ 1 + \frac {\delta \chi}{\chi_0}\right\} \bar \psi \psi,
\ee
where we used that the scaling dimension of a fermionic bilinear operator $\bar \psi \psi$ is 3,
the dimension of $d^4 x$ in the action is $-4$ and hence one dilaton insertion is needed to recover scale invariance. 
Furthermore, additional interactions can be generated by the sources of explicit scale invariance breaking in the conformal sector. For example the fermionic mass $m_\psi$ can be proportional to a Yukawa coupling $\lambda_\psi$ which runs with energy.
At the condensation scale $\chi$ the new-physics degrees of freedom other than the dilaton should become heavy and hence decouple from the RG evolution. To account for their $\chi$-dependent contribution in the running till that scale we should then use
\be
\lambda_\psi \; \;  \to \; \;  \lambda_\psi \left(1 + \left[\frac{\partial \log \lambda_\psi}{\partial \log \mu}\right]_{\text{CFT}} \frac{\delta \chi}{\chi_0}\right),
\ee
where the term in square brackets only includes the contribution to the running from the decoupled conformal field theory (CFT) degrees of freedom.
This introduces the following additional contribution to the dilaton interaction
\be
{\cal L}_{m_\psi} \; \to \; m_\psi \left\{ 1+ \frac {\delta \chi}{\chi_0} + \left[\frac {\partial \log \lambda_\psi}{\partial \log \mu}\right]_{\text{CFT}} \frac{\delta \chi}{\chi_0} \right\} \bar \psi \psi + \dots
\ee
Finally, the dilaton couplings can be altered in the presence of mass mixing between the dilaton and the Higgs boson. The diagonalisation of the Higgs-dilaton mass matrix amounts to a redefinition  
\be\label{eq:massbasis}
\chi = \chi_0+ c_\theta \hat \chi - s_\theta \hat h, \quad \;  \; h = v + c_\theta \hat h + s_\theta \hat \chi\,,
\ee 
where $\hat h$ and $\hat \chi$ are the mass eigenstates.
Assuming that the mass of the fermion $\psi$ scales like (as typically the case in Composite Higgs models)
\be
 m_\psi \propto \sin (h/f) 
 \ee
 with  $f$ being the Higgs decay constant (more on that in the next sections), we find that the leading interaction of the dilaton-like mass eigenstate is given by
\be
 {\cal L}_{m_\psi} \; \; \to \; \; {m_\psi} \bar \psi \psi \hat \chi \left\{s_\theta \frac 1 v + c_\theta \frac 1 {\chi_0} \left(1+ \left[\frac {\partial \log \lambda_\psi}{\partial \log \mu}\right]_{\text{CFT}} \right)  \right\}    + \dots\, ,
\ee
where we neglected $(v/f)^2$ corrections.
For the pseudo-Goldstone Higgs boson the mixing angle $\theta$ is vanishing in the limit of exact scale invariance and hence is sensitive to the explicit breaking of the latter~\cite{Chacko:2012sy,Bruggisser:2018mrt}. We will discuss the mixing in more detail Section~\ref{sec:mixing}.

Following these simple rules, in Section~\ref{sec:colliderpheno} we will write down explicitly the relevant dilaton couplings to SM states.

\subsection{Connection to the Higgs Potential and EW Scale Naturalness}

We have discussed generic features of the dilaton interactions with SM states external to the CFT sector. We will now assume that the Higgs boson is a composite state produced by the CFT dynamics. To ensure its lightness compared to other CFT states with mass $m_{\rm CFT}\propto \chi_0$ we will consider the case that the Higgs is a pseudo Nambu-Goldstone boson (NGB) of some approximate flavour symmetry $G$ of the CFT sector which is spontaneously broken to a subgroup $H$ at the CFT condensation scale. For example the minimal $SO(5) \to SO(4)$ breaking pattern~\cite{Agashe:2004rs}, with the SM $SU(2)_L$ embedded into $SO(5)$, gives exactly four NGBs allowing to form the complex Higgs doublet. The Higgs compositeness which we assume here allows to address the EW scale naturalness problem, up to a relatively small residual fine-tuning~\cite{Grojean:2013qca}. In this section we will discuss how this fine-tuning is related to the dilaton couplings. 

The Higgs potential is generated by the couplings which explicitly break the $G$ invariance, and include the SM gauge couplings and the top quark Yukawa coupling.\footnote{The EW symmetry group is embedded into $G$ but the SM states only form incomplete $G$-multiplets, hence their interactions break $G$.} The resulting one-loop scalar potential has the following form at the leading order in elementary-composite interactions~\cite{Matsedonskyi:2012ym,Panico:2015jxa}
\be\label{eq:vCHtree}
V_h = \alpha \sin^2 h/f + \beta \sin^4 h/f,
\ee
where $f$ is the scale of $G \to H$ symmetry breaking. As we will discuss later, this scale can be parametrically different from the scale $\chi_0$. By expanding the trigonometric functions in the above formula, one sees that $f$ controls higher-order interactions of the Higgs boson, which are absent in the SM. Analogous higher-order interactions appear in all Higgs couplings, and are constrained by experimental data. Currently, the parameter $\xi \equiv (246\text{ GeV}/f)^2$ is restricted to be around or less than $0.1$~\cite{Grojean:2013qca}, or, equivalently 
\be
f \gtrsim 800 \mbox{ GeV}.
\ee 
At the same time, the Higgs VEV is dictated by Eq.~(\ref{eq:vCHtree}) and reads $h^2=-(1/2) (\alpha/\beta) f^2$. The Higgs VEV is then expected to be of order $f$, unless $\alpha$ and $\beta$ are finely tuned, since one generically expects $\alpha \gtrsim \beta$ in this type of theories.\footnote{See e.g.~\cite{Durieux:2021riy} for a proposal which could remove this tuning.} The corresponding degree of unnatural fine-tuning is given by $\xi$: weaker tuning requires larger $\xi$ and lower $f$. The parameter $f$ hence has a paramount importance for Higgs physics and EW scale naturalness. We will now discuss how it is connected to the dilaton scale $\chi_0$.

To gain parametric estimates of the properties of the conformal sector we assume that it behaves as an $SU(N)$ confining QCD-like theory. 
The parameter $f$ corresponds to the VEV of some condensate transforming non-trivially under the global symmetry group $G$ and breaking the latter spontaneously. We will therefore associate $f$ with an analogue of the quark-antiquark condensate, with its excitations (such as the Higgs) corresponding to meson-like states. 
The dilaton scale $\chi$, on the other hand, corresponds to the VEV of a condensate which can a priori be neutral under $G$, and whose excitations can be either glueball-like or meson-like states. Although in the analyses based on the AdS/CFT correspondence the dilaton follows the glueball-like behaviour, we retain the meson-like option to be able to capture a  more general class of theories. As a matter of fact, most of the previous studies of dilaton phenomenology in the composite Higgs context were equating $f$ and $\chi_0$ which, as we discuss below, corresponds to a meson-like $\chi$. 

According to Ref.~\cite{Witten:1979kh}, in an $SU(N)$ theory, each canonically normalised meson field enters the Lagrangian with a factor of $1/\sqrt N$, while each glueball field is accompanied by $1/N$. 
This can be reflected by defining the following couplings for mesons and glueballs 
\be\label{eq:largencoupl}
g_{\text{mes}} = 4 \pi/ \sqrt{N}, \quad \; \;
g_{\text{glue}} = 4 \pi/ N,
\ee
where the factors of $4 \pi$ are chosen to reproduce a fully strongly coupled theory as $N\to1$. At the same time, Ref.~\cite{Witten:1979kh} shows that the masses of meson- and glueball-like states do not scale with $N$. Using dimensional analysis one can then recover the scaling of the VEVs $\langle ... \rangle$ of the different condensates with $N$:
\be
\langle \text{meson} \rangle \propto \frac{m_{\text{mes}}}{g_{\text{mes}}} \propto \sqrt{N},\quad \;\;
\langle \text{glueball} \rangle \propto \frac{m_{\text{glue}}}{g_{\text{glue}}} \propto N.
\ee
This implies the following $N$-scaling relation between the decay constants of the Higgs and the dilaton
\be\label{eq:fchi0}
f  \propto \frac {g_\chi}{g_{\text{mes}}} \chi_0 ,
\ee
where $g_\chi=g_{\text{mes}}(g_{\text{glue}})$ for a meson (glueball) dilaton
(see~\cite{Bruggisser:2018mrt} for an alternative derivation of this $N$-scaling). In the case of a glueball-like dilaton this becomes
\be
\label{eq:fchi0glu}
 f  \propto {\chi_0}/{\sqrt{N}} \quad \mbox{(glueball-like dilaton)},
\ee
allowing for $\chi_0$ to be much larger than $f$ for large $N$. For a meson-like dilaton, on the other hand, we find 
\be
 f  \propto \chi_0 \quad \mbox{(meson-like dilaton)}, 
\ee
independent of $N$.

An intuitive way (although eventually based on the same large-$N$ counting) to derive the $\sqrt N$-enhancement of the gluon condensate~(\ref{eq:fchi0glu}) is the following. Let us assume that the conformal symmetry breaking is driven by the gluon condensate $\chi$, which also interacts with a quark-antiquark condensate $\sigma$ and sets the scale of the latter. According to the large-$N$ counting~\cite{Witten:1979kh}, the glueball-meson scattering amplitude is $1/N^2$ suppressed, while the meson-meson scattering amplitude goes like $1/N$. The corresponding potential would schematically look like
\be
V_\sigma= - \frac 1 {N^2} \chi^2 |\sigma|^2 +  \frac 1{N} |\sigma|^4,
\ee
which leads precisely to the dependence of $f$ on the dilaton VEV derived above: $f =\langle\sigma\rangle \sim \langle\chi\rangle/\sqrt{N}$. 

Finally, an analogous relation between $\chi_0$ and $f$ can also be derived based on the AdS/CFT correspondence using 5D dual models.
In this case the NGB composite Higgs can be modelled by the 5th component of a gauge field propagating in the bulk. The Higgs decay constant $f_\text{RS}$ is then given by~\cite{Agashe:2004rs}\footnote{We skip a proper introduction to the higher-dimensional dual theories which is beyond the scope of this paper, but can be found in numerous reviews.}
\be
f_\text{RS}\,=\, \mu_{\text{RS}} \frac 2 {g_5 \sqrt k} ,
\ee 
where $\mu_{\text{RS}}$ is the VEV of the radion and $g_5 \sqrt k$ approximately equals the coupling of KK modes of the bulk gauge field~\cite{Contino:2003ve}. These KK excitations should correspond to meson-like composite states in the 4D dual theory, and hence we fix $g_5 \sqrt k=g_{\text{mes}}$. Assuming that this coupling follows the large-$N$ estimate, $g_\text{mes} = 4 \pi/\sqrt N$, one obtains the relation $f_\text{RS} \propto \mu_{\text{RS}} \sqrt N$. This scaling can also be derived by an explicit string theory computation, see e.g.~\cite{Sakai:2004cn}.
Now, switching to the canonically normalized radion $\tilde \mu_{\text{RS}} = \sqrt{24} \mu_{\text{RS}}/g_\chi$, dual to the dilaton $\chi$, with $g_\chi = 4 \pi/N$ fixed by AdS/CFT, we get
\be
f_\text{RS}\,=\, \tilde \mu_{\text{RS}} \frac {2}{\sqrt{24}} \frac {g_\chi} {g_{\text{mes}}} \,\simeq\, \tilde \mu_{\text{RS}} \frac {0.4}{\sqrt{N}}. 
\ee 
This has the same $N$-dependence as was obtained in Eq.~(\ref{eq:fchi0glu}), up to an overall order-one factor.
Generally speaking, the large-$N$ expansion allows to estimate the size of various quantities only up to order-a-few factors. To account for this ambiguity we introduce a coefficient $c_{h\chi}$ in Eq.~(\ref{eq:fchi0glu}), such that the relation between $\chi_0$ and $f$ reads
\be
\label{eq:twoscales}
\chi_0=c_{h \chi}\left(\frac{g_{\text{mes}}}{g_\chi}\right) f= c_{h \chi} f 
 \begin{cases}
\sqrt{N} & \; \; \mbox{for glueball-like dilaton}\\
1 & \; \; \mbox{for meson-like dilaton.}\\
\end{cases}
\ee
In the following we will let $c_{h \chi}$ vary within an order-one range.

\subsection{Dilaton-Higgs Mixing and Minimal Mass Splitting}\label{sec:mixing}

The dilaton and Higgs phenomenology depends significantly on the mass mixing between the two fields, as we will discuss in detail in the next sections.
This mass mixing can only be induced by operators which include both sources breaking the Higgs shift symmetry and those breaking conformal invariance~\cite{Chacko:2012sy,Bruggisser:2018mrt}. Indeed, in the absence of Higgs-shift-symmetry breaking the Goldstone Higgs can not have any potential at all. While in the absence of conformal-invariance breaking the Higgs-dependent part of the potential reads $V_{h\chi} \propto \chi^4 V(h/f)$, with $\partial_h V(h/f) = 0$ in the minimum of the scalar potential. Hence the mass mixing $\partial_h \partial_\chi V_{h\chi}$ vanishes too.

The main source of Higgs-shift-symmetry breaking is the top quark Yukawa coupling, while for the breaking of conformal invariance we can single out two distinct types of sources. The first one is the breaking induced by 
interactions of the (nearly) conformal sector, and the second type comes from the breaking induced by coupling the 
conformal sector
with external elementary fields. The main practical difference for the purpose of our discussion is that the latter type can {\it a priori} be unrelated to the dilaton mass, as we discuss in the following.

We will use the following estimate for the mass mixing
\be\label{eq:massmix}
m_{h \chi}^2 \simeq m_*^2 \times \left[\frac{3 \lambda_t^2}{16 \pi^2}\right] \times \left[ \gamma_{\text{comp}} + \gamma_{\text{elem}}\right] \times \left[\frac{f}{\chi_0}\right].
\ee
The factor 
\be
m_*^2 = g_{\text{mes}}^2 f^2 = g_\chi^2 \chi_0^2/ c_{h \chi}^2
\ee
 is a generic coefficient of mass-dimension-two operators generated by the composite sector~\cite{Witten:1979kh} if no selection rules or symmetry suppression apply. The first square brackets in (\ref{eq:massmix}) contain the estimated size of the Higgs-shift-symmetry breaking induced by a loop with an elementary top quark. The two parameters in the second square brackets, $\gamma_{\text{comp}}$ and $\gamma_{\text{elem}}$, parametrise the conformal-invariance breaking induced by the nearly-conformal sector itself, and by its interactions with the elementary states, respectively. 

The parameter $\gamma_{\text{comp}}$ is related to the dilaton mass. Indeed, the general form of the dilaton potential in the presence of a scalar CFT operator ${\cal O}_\epsilon$ with a running coefficient $\epsilon(\mu)$ is
\be
V_\chi = g_\chi^2 \chi^4 + g_\chi^2 \epsilon(\chi) \chi^4.
\ee
This results in the dilaton mass $m_\chi^2 \sim \gamma_{\text{comp}} m_*^2$~\cite{Bruggisser:2018mrt}, where $\gamma_{\text{comp}}  = \gamma_\epsilon = \partial \log \epsilon /\partial \log \mu$. Hence an insertion of a conformal-invariance-breaking operator ${\cal O}_\epsilon$ in a loop diagram produces a factor $\epsilon (\chi) \simeq \epsilon (\chi_0)(1+\gamma_\epsilon (\chi-\chi_0)/\chi_0)$. The non-trivial dependence on the dilaton thus comes with a factor $\gamma_\epsilon \sim m_\chi^2/m_*^2$.

The second source of conformal-invariance breaking, corresponding to $\gamma_{\text{elem}}$, is assumed to be generated by interactions with the elementary fields. For example, if the top quark Yukawa coupling varies significantly with the dilaton VEV, 
we obtain $\gamma_{\text{elem}} \sim \partial \log\lambda_t /\partial \log \mu$. Note that, unlike $\gamma_{\text{comp}}$,  $\gamma_{\text{elem}}$ can only contribute to the dilaton mass due to loops with elementary fermions, hence the contribution of $\gamma_{\text{elem}}$ to $m_\chi^2$ has to be suppressed by a loop factor. We therefore assume that the dilaton mass is mostly determined by $\gamma_{\text{comp}}$, while $\gamma_{\text{elem}}$ is not constrained and can, in particular, be greater than $\gamma_{\text{comp}}$ and give the main contribution to the mass mixing term $m_{h \chi}^2$. This is expected to happen for example in the models of Ref.~\cite{Bruggisser:2018mrt,Bruggisser:2018mus,vonHarling:2016vhf}.

Finally, the factor ${f}/{\chi_0}$ in Eq.~(\ref{eq:massmix}) can be deduced from dimensional analysis~\cite{Bruggisser:2018mrt}, and in the case of a glueball-like dilaton represents the expected $1/\sqrt N$-suppression of the glueball-meson mixing.  

Let us now discuss the effect of the mass mixing term on the mass diagonalization. First of all, diagonalizing the mass matrix $\{\{ m_{hh}^2, m_{h\chi}^2 \},\{m_{h\chi}^2, m_{\chi \chi}^2 \} \} \to \text{diag}\{ m_h^2, m_\chi^2 \}$ we see that the dilaton-Higgs mass splitting should satisfy
\be\label{eq:minsplit}
|m_\chi^2 - m_h^2| \geq 2 |m_{h \chi}^2|,
\ee
otherwise obtaining the desired dilaton mass is not possible. 
Furthermore, the mixing angle between the dilaton and the Higgs is given by~\cite{Fuchs:2020cmm}
\be
|\sin \theta| = \frac 1 {\sqrt{2}} \sqrt{1- \sqrt{1- 4 \frac{m_{h \chi}^4}{(m_\chi^2-m_h^2)^2}}} 
\simeq \left|\frac{m_{h \chi}^2}{m_\chi^2-m_h^2} \right|,
\ee 
where in the last step we have expanded for small mass mixing. The mixing angle grows as the mass difference $|m_\chi^2 - m_h^2|$ approaches $2 |m_{h \chi}^2|$. Hence generally, the constraints on the deviations of the Higgs  couplings from the SM, which are sensitive to $\sin \theta$ (see Section~\ref{sec:hcoupl}), are expected to exclude a larger part of the parameter space than the minimal splitting condition.  
As we will show in the following, a moderate value of $\gamma_{\text{elem}}$ excludes a significant fraction of the model parameter space at low dilaton masses.

\section{Dilaton Phenomenology}\label{sec:colliderpheno}

Let us now write down explicitly all the relevant dilaton couplings to the SM states.
To fix the conventions, we parametrise the transformation to the mass eigenstates $\hat h, \hat \chi$ by
\be\label{eq:massbasis}
\chi = \chi_0+ c_\theta \hat \chi - s_\theta \hat h, \quad \; \; h = v_{\text{CH}} + c_\theta \hat h + s_\theta \hat \chi\,,
\ee 
where $s_\theta, c_\theta$ are the sine and cosine of the mixing angle and 
\be\label{eq:hvevs}
\sin(v_{\text{CH}}/f) = v_{\text{SM}}/f
\ee
with $v_{\text{SM}}=246$~GeV. The relation in Eq.~(\ref{eq:hvevs}) is due to the fact that the NGB Higgs enters the Lagrangian in the form of trigonometric functions such that e.g.~$m_W = g f \sin (v_{\text{CH}}/f)/2$. 

To evaluate the LHC bounds on the dilaton we will use the dilaton couplings and the partial widths detailed below, derived using the arguments presented in the previous section.

\subsection{Fermions}

For definiteness we will assume the following form for the fermionic mass terms
\be
{\cal L} \supset -\frac {\lambda_\psi}{\sqrt 2} (\chi/\chi_0) f \, \sin(h/f) \bar \psi \psi,
\ee
although the trigonometric dependence on the Higgs VEV can take different forms, depending on how the elementary fermions are embedded into the group $G$. The Yukawa coupling $\lambda_\psi$ can significantly run with energy in the CFT regime and we define~\footnote{In the simplest case of partial compositeness~\cite{Kaplan:1991dc} the Yukawa interactions are proportional to the product of the mass mixing between the composite states with the left- and right-handed elementary fermions, $y_{L,R}$, so that $\gamma_{\psi} = \gamma_{y_L} + \gamma_{y_R}$.}
\be
\gamma_\psi = [\partial \log \lambda_\psi/\partial \log \mu]_{\text{CFT}}.
\ee
The expression for fermion-dilaton interactions was derived in Section~\ref{sec:coupling} and we only make it more precise by adding the terms subleading in $v^2/f^2$:
\bea
{\cal L}& 
\supset& 
-\sum_\psi \frac {\lambda_\psi} {\sqrt 2} \left\{s_\theta \sqrt{1-v_{\text{SM}}^2/f^2}+ c_\theta (1+\gamma_{\psi}) \frac{v_{\text{SM}}}{\chi_0}  \right\} \bar \psi \psi \hat \chi + \text{h.c.} \\
&\equiv&-\sum_\psi \frac {\lambda_\psi} {\sqrt 2} \, \kappa_{\psi}^\chi \, \bar \psi \psi \hat\chi + \text{h.c.}  \label{eq:Lchiferm}
\eea
Here the presence of a small $v^2/f^2$ correction is specific to the NGB Higgs, while the other terms are generically expected for the dilaton interactions.
The parameter $\kappa_{\psi}^\chi$ defined in the last line is the ratio of the dilaton-fermion coupling to the SM Higgs-fermion coupling. 
The corresponding dilaton decay width at  leading order is given by~\cite{Spira:2016ztx}
\bea
\Gamma^\chi_\psi = (\kappa_{\psi}^\chi)^2 \frac {N_{c\psi} G_F m_\chi}{4\sqrt 2 \pi} m_\psi^2 \left(1-4\frac {m_\psi^2}{m_\chi^2} \right)^{3/2},
\eea
where $N_{c\psi}=1, 3$ for leptons and quarks, respectively, and $G_F=1/\sqrt 2 v_{\text{SM}}^2$. 

\subsection{Massive Vectors}

The mass of the $W$ boson is given by
\be\label{eq:wmass}
{\cal L} \supset \frac{g^2} 4 f^2 \sin^2(h/f) \, (\chi^2/\chi_0^2) \, |W_\mu|^2,
\ee
and similarly for the $Z$. 
Applying the sequence of derivations from Section~\ref{sec:coupling} to Eq.~(\ref{eq:wmass}) we obtain the dilaton interactions
\bea
\label{eq:chivvmass}
{\cal L} & \supset&
2 \frac{\hat\chi}{v_{\text{SM}}}  \left\{{s_\theta} \sqrt{1-v_{\text{SM}}^2/f^2} + c_\theta \left(1+ \gamma_{V^2} \right) \frac {v_{\text{SM}}}{\chi_0} \right\} \left(m_W^2 |W_\mu|^2 + \frac 1 2 m_Z^2 Z_\mu^2 \right) \\
& \equiv&2 \frac{\hat\chi}{v_{\text{SM}}}  \,\kappa_{V}^\chi \, \left(m_W^2 |W_\mu|^2 + \frac 1 2 m_Z^2 Z_\mu^2 \right),
\label{eq:chivvmass2}
\eea
where $\kappa_{V}^\chi$ is the ratio of the dilaton and SM Higgs couplings to massive vectors. The quantity $\gamma_{V^2}$ parametrises a possible scale-invariance-breaking contribution 
of the nearly-conformal sector
to the renormalisation of the $h^2 W^2$ operator.

Additionally, the kinetic terms of the EW gauge bosons,
\be
{\cal L} \supset - \frac 1 4 W^a_{\mu \nu} W^{a \mu \nu} - \frac 1 4 B_{\mu \nu} B^{\mu \nu},
\ee
can also give rise to interactions with the dilaton. To derive them, it is convenient to first switch to non-canonical gauge fields with kinetic terms
\be
{\cal L} \supset  - \frac 1 4 \frac 1 {g^2} W^a_{\mu \nu} W_{a\mu \nu} \to -\frac 1 4 \frac 1 {g^2} \left(1 - 2 \frac 1 g \frac{\partial g}{\partial \log \mu} \frac {\delta \mu}{\mu} \right) W^a_{\mu \nu} W^{a \mu \nu}.
\ee
In the second step we expressed the coupling in the first step which is renormalised at some scale $\mu'$ 
as a coupling renormalised at $\mu+\delta \mu$.  
Next one notices that the renormalisation of the $W_{\mu \nu}^2$ operator is the only contribution to the gauge coupling running $\partial g / \partial \log \mu$ which is sensitive to EW-charged CFT states with masses $\sim \chi$ after confinement.
Then the dependence of $W_{\mu \nu}^2$ on the dilaton (unlike e.g.~that of the operator in Eq.~(\ref{eq:wmass})) has to be captured by the contribution of 
these states to the $\beta$-function. 
Setting $\delta \mu$ to $\delta \chi$ and $\mu$ to $\chi_0$, and switching back to canonically normalised fields we obtain
\be\label{eq:Wkinetic}
{\cal L} \supset \frac 1 2 W^a_{\mu \nu} W_{a\mu \nu} \frac{1}{g}\left[ \frac{\partial g}{\partial \log \mu}  \right]_{\text{CFT}} \frac{c_\theta \hat \chi}{\chi_0}.
\ee
We are interested here in the contribution of the heavy CFT states to the coupling, while the contribution of light states present in our EFT 
will be evaluated explicitly as loop corrections. The quantity in square brackets 
is therefore the jump of the $\beta$-function induced by the decoupling of EW-charged CFT states which become heavy due to confinement. One expects that this jump is proportional to the number of colors $N$~\cite{Garriga:2002vf}:
\be\label{eq:betaW}
[{\partial \log g}/{\partial \log \mu}]_{\text{CFT}} \sim N (g^2/16\pi^2) =  g^2/g_{\text{mes}}^2.
\ee
However, the change of the running below and above the confinement scale depends on whether some CFT states remain light below that scale. This could happen if some of the SM states (e.g.~the right-handed top, the Higgs boson or the longitudinal components of the gauge fields) are completely composite. If these states contribute to the $\beta$-function below $\chi_0$, this contribution has to be subtracted from the r.h.s.~of~(\ref{eq:betaW}). The multiplicity of the light states is however not expected to scale with $N$. Hence in the large-$N$ limit this correction is subdominant.  
For the following we will define 
\be
[{\partial \log g}/{\partial \log \mu}]_{\text{CFT}} = (c_{WW} - \tilde c_{WW}/N)  \, g^2/g_{\text{mes}}^2 .
\ee 
The coefficients $c_{WW}$ and $\tilde c_{WW}$ can vanish or be of order a few, with $\tilde c_{WW}$ corresponding to the subtracted contribution of light composite states discussed above. Analogously, we define the jump of $\beta_{g'}/g'$ determining the coupling to the vector boson $B_\mu$  as 
\be
[{\partial \log g'}/{\partial \log \mu}]_{\text{CFT}} = (c_{WW} - \tilde c_{BB}/N) \,  g'^2/g_{\text{mes}}^2. 
\ee

With all the discussed interactions included, the dilaton decay widths to on-shell gauge bosons are given by~\cite{Ahmed:2019csf}
\begin{multline}
\Gamma^\chi_{V} =g_* \frac {G_F m_\chi^3}{16\sqrt 2 \pi} \sqrt{1-4\frac {m_{V}^2}{m_\chi^2} } \left((\kappa_{V}^{\chi2}+8\kappa_{VV}^{\chi2})\right.  \\ \left. -4 (\kappa_{V}^{\chi 2}+6\kappa_{V}^{\chi} \kappa_{VV}^{\chi} + 8 \kappa_{VV}^{\chi2}) \frac {m_V^2}{m_\chi^2} +12 (\kappa_{V}^\chi+2 \kappa_{VV}^{\chi})^2 \frac {m_V^4}{m_\chi^4} \right), 
\end{multline}
where $V={\{Z,W\}}$ and $g_*={\{1,2\}}$ respectively. 
Furthermore, the $\kappa_{V}^{\chi}$ are defined in Eq.~\eqref{eq:chivvmass2} and the $\kappa_{VV}^{\chi}$ follow from  the interactions of the type in Eq.~\eqref{eq:Wkinetic} and read
\bea
\kappa_{WW}^{\chi} & = &\frac 1 2  \frac{c_\theta v_{\rm SM}}{\chi_0} \frac{g^2}{g_{\text{mes}}^2} c_{WW}\\
\kappa_{ZZ}^{\chi}  & = & \frac 1 2 \frac{c_\theta v_{\rm SM}}{\chi_0} \left\{ c_w^2 c_{WW} \frac{g^2}{g_{\text{mes}}^2} +   s_w^2 c_{BB} \frac{g^{\prime 2}}{g_{\text{mes}}^2} \right\} 
\eea
with $s_w, c_w$ being the sine and cosine of the weak angle respectively. We omitted small corrections due to the SM loops which are not enhanced by the potentially large $N$ and hence are always subdominant compared to the tree-level coupling in Eq.~(\ref{eq:chivvmass}). The same applies to the contributions $\propto \tilde c_{BB,WW}$.

\subsection{Gluons}

The dilaton can acquire couplings to gluons through loop diagrams with quarks in the same way as the Higgs does. The dilaton couplings to the quarks in turn can be simply read off from Eq.~(\ref{eq:Lchiferm}). The resulting dilaton decay width is given by
\bea\label{eq:gammagg1}
\Gamma^\chi_{g} =\frac {G_F \alpha_s^2 m_\chi^3}{36\sqrt 2 \pi^3} \Big|\sum_q \kappa_{q}^\chi  A_q (\tau_q)\Big|^2, 
\eea
where $\tau_q = 4 m_q^2/m_\chi^2$ and the loop function $A_q(\tau)$, with $A_q(\infty)\to 1$,  is given in Ref.~\cite{Spira:2016ztx} and also quoted in Appendix~\ref{sec:loopfunctions} for completeness.

Additionally, interactions between the dilaton and the gluons can be induced by the heavy CFT states. These can be related to the running of the gluon gauge coupling $g_s$ induced by QCD-charged CFT states, analogous to the running of the EW couplings discussed in the previous section. The resulting dilaton coupling is 
\be
{\cal L}\supset \frac 1 2 \left[ \frac{\partial \log g_s}{\partial \log \mu}  \right]_{\text{CFT}} G_{\mu \nu}^a G^{a \mu \nu} \frac{c_\theta \hat\chi}{\chi_0},
\ee
where we define $[{\partial \log g_s}/{\partial \log \mu}]_{\text{CFT}} \equiv (2/3) (c_{GG}-\tilde c_{GG}/N) g_s^2/g_{\text{mes}}^2$ such that for $c_{GG}=1$ each unit of $N$ contributes to the $\beta$-function as a single SM quark flavour.
The dilaton decay width can be obtained from the previous result~(\ref{eq:gammagg1}) by the substitution
\be
\sum_q \kappa_{q}^\chi  A_q (\tau_q) \to \sum_q \kappa_{q}^\chi  A_q (\tau_q) + c_{GG}\, c_\theta\, N \frac{v_{\text{SM}}}{\chi_0},
\ee
where we again neglected $\tilde c_{GG}$. 
The dilaton-gluon coupling modifier (with respect to the gluon coupling to a SM-like Higgs boson with mass $m_\chi$) is therefore 
\bea
\kappa_g^\chi = \Big|\sum_q \kappa_{q}^\chi  A_q^\chi (\tau_q)+ c_{GG}\, c_\theta\, N \frac{v_{\text{SM}}}{\chi_0}\Big|/\Big|\sum_q A_q^\chi (\tau_q)\Big|.
\eea

\subsection{Photons}

The situation here is analogous to the case with gluons.  The resulting photon contribution to the dilaton decay width can be parametrized as
\bea
\Gamma^\chi_{\gamma} =\frac {G_F \alpha^2 m_\chi^3}{128\sqrt 2 \pi^3} \Big|\sum_f \kappa_{f}^\chi N_{cf} e_f^2  A_f (\tau_f) + \kappa_{W}^\chi A_W (\tau_W) + c_{\gamma\gamma}\, c_\theta\, N \frac{v_{\text{SM}}}{\chi_0} \Big|^2, 
\eea
where $c_{\gamma\gamma} = (3/4)(c_{WW} +c_{BB})$ corresponds to the size of the heavy new-physics contribution,
such that 
$[{\partial \log e}/{\partial \log \mu}]_{\text{CFT}} \equiv (4/3) c_{\gamma \gamma} e^2/g_{\text{mes}}^2$.
The loop functions $A_f(\tau)$ and $A_W(\tau)$ can again be found in Appendix~\ref{sec:loopfunctions}.
The dilaton-photon coupling modifier is
\bea
\kappa_{\gamma}^\chi = \Big|\sum_f \kappa_{f}^\chi N_{cf} e_f^2  A_f (\tau_f) + \kappa_{W}^\chi A_W (\tau_W)+ c_{\gamma\gamma}\, c_\theta\, N \frac{v_{\text{SM}}}{\chi_0}\Big| / \Big|\sum_f N_{cf} e_f^2  A_f (\tau_f) + A_W (\tau_W) \Big|. \nn \\
\eea

\subsection{$Z \gamma$}

The SM particles can mediate a $h Z \gamma$ interaction at the one-loop level. Analogously, the dilaton acquires a $\chi Z \gamma$ interaction, whose strength is modified with respect to that of the Higgs due to the modified dilaton couplings to SM fermions and vectors. The width of the one-loop induced dilaton decay into $Z \gamma$ is then given by
\be\label{eq:chigaz}
\Gamma^\chi_{Z\gamma}= \frac{G_F^2 M_W^2 \alpha m_\chi^3}{64 \pi^4} \left(1-\frac{M_Z^2}{m_\chi^2}\right)^3 \left| \sum_f \kappa_f^\chi A_f [\tau_f, \sigma_f] + \kappa_W^\chi A_W[\tau_W, \sigma_W] \right|^2
\ee
with $\tau_i = 4 m_i^2/m_\chi^2$ and $\sigma_i = 4 m_i^2/m_Z^2$.  The functions $A_f[\tau,\sigma]$ and $A_W[\tau,\sigma]$ are listed in Appendix~\ref{sec:loopfunctions}.

Additionally, there can be pure CFT contributions to the considered process. The main effect growing with $N$ comes from the renormalisation of the couplings  $g$ and $g'$ which generate the operator
\be
{\cal L} \supset \frac {c_\theta \hat \chi}{\chi_0} s_w c_w \left\{ c_{WW} \frac{g^2}{g_{\text{mes}}^2} - c_{BB} \frac{g^{\prime2}}{g_{\text{mes}}^2}  \right\}  Z_{\mu \nu} \gamma^{\mu \nu}.
\ee
To incorporate the corresponding contribution into the decay width~(\ref{eq:chigaz}) one should perform a shift~\cite{Dawson:2018pyl}
\be
\sum_f \kappa_f^\chi A_f^{\chi} [\tau_f, \sigma_f] + \kappa_W^\chi A_W^\chi[\tau_W, \sigma_W] \to \sum_f \kappa_f^\chi A_f^{\chi} [\tau_f, \sigma_f] + \kappa_W^\chi A_W^\chi[\tau_W, \sigma_W] - \kappa_{Z \gamma}^{\chi},
\ee
where 
\be
\kappa_{Z \gamma}^{\chi}= \frac{16\pi^2}{g_{\text{mes}}^2} \frac{c_\theta v_{\rm SM}}{\chi_0} s_w \left\{ (c_w/s_w) c_{WW} - (s_w/c_w) c_{BB} \right\}.
\ee

\subsection{Higgs}

The relevant Higgs-dilaton interactions are given by (neglecting possible scale-invariance breaking)
\bea
{\cal L} \supset - \left(2\frac{m_h^2}{\chi_0}  \right)\hat \chi \hat h^2 
\equiv a_{h \chi} \hat \chi \hat h^2,
\eea
and contribute to the dilaton decay width with
\be
\Gamma^\chi_h= \frac{a_{h \chi}^2}{8 \pi m_\chi} \left(1-4\frac {m_h^2}{m_\chi^2} \right)^{1/2} \Theta(m_\chi-2 m_h).
\ee

\subsection{Higgs-Coupling Modifications}\label{sec:hcoupl}

For completeness we should mention that the dilaton-Higgs mixing angle which affects the dilaton phenomenology can be constrained from Higgs physics, although we will not analyse that in this work in much detail. The least model-dependent constraint along these lines comes from the modification of the Higgs couplings to the EW gauge bosons. The corresponding coupling modifier with respect to the SM prediction reads
\be\label{eq:kapavh}
\kappa_{V}^{h} = c_\theta \cos \frac{v_{\rm CH}}{f}  -  s_\theta \frac{g_\chi}{g_*} (1+\gamma_{V^2}) \sin \frac{v_{\rm  CH}}{f},
\ee
which can be derived from the expression for the $W$ mass~(\ref{eq:wmass}). In the limit of $g_\chi=g_*$ and $\gamma_{V^2}=0$ this expression simplifies to
\be
\kappa_{V}^{h} = \cos \left(\theta+ \frac{v_{\rm  CH}}{f} \right).
\ee

\begin{figure}[t]
\centering
\includegraphics[width=5.cm]{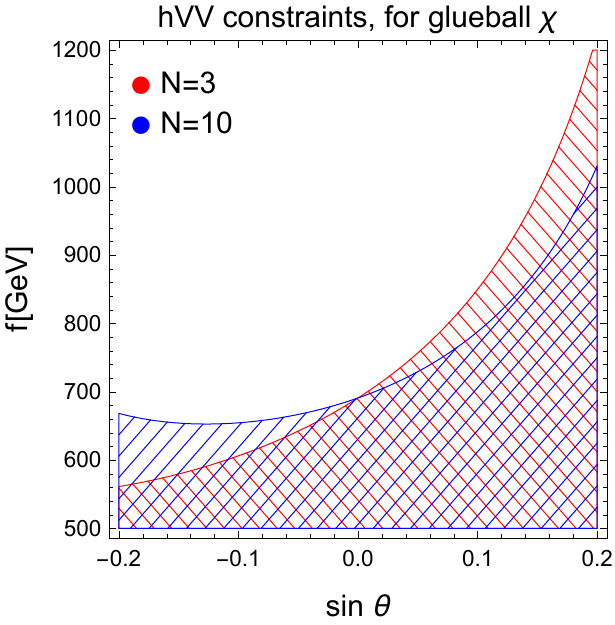}
\hspace{1.5cm}
\includegraphics[width=5.cm]{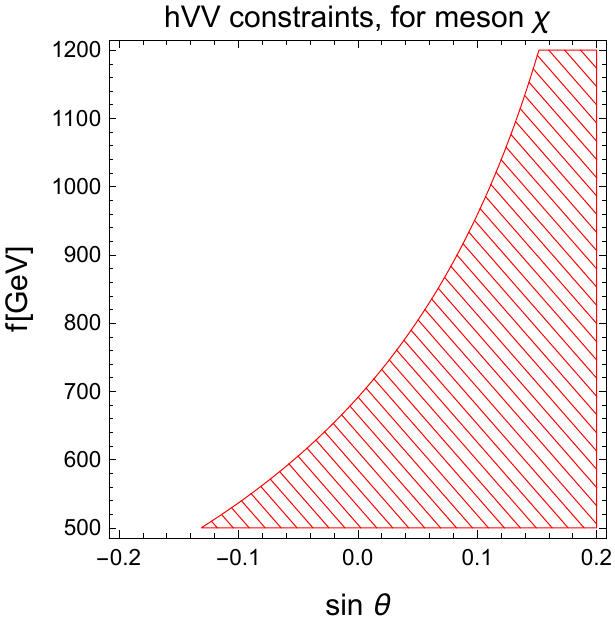} 
\caption{\small \it{Current bounds on the dilaton-Higgs mixing angle and $f$ derived from the Higgs-EW vector boson coupling measurements.}}
\label{fig:plothVV}
\end{figure}

Therefore, if $\theta$ is negative, it can compensate the Higgs coupling distortion introduced by non-zero $v/f$, thus bringing the couplings closer to their SM values. On the other hand, the result~(\ref{eq:kapavh}) can be interpreted as the possibility to access the degree of conformal-invariance breaking in the UV by measuring the Higgs couplings. Using the currently available constraints on the Higgs-vector boson coupling modifications from direct measurements~\cite{CMS:2022dwd,ATLAS:2022vkf} we present the $2 \sigma$ bounds on the Higgs-dilaton mixing angle and the scale $f$ in Fig.~\ref{fig:plothVV}, for $\gamma_{V^2}=0$ and $c_{h \chi}=1$. Note that $\sin \theta$ and $f$ can also be constrained from other measurements, whose detailed analysis would however bring us outside the scope of this paper.

\section{Collider Bounds}\label{sec:colliderbounds}

In this section, we present the current 95\%CL LHC exclusion limits for the parameter space of the dilaton EFT, as well as the projected future HL-LHC sensitivity. We derive the bounds using HiggsTools and related software packages~\cite{Djouadi:2018xqq,Bechtle:2020pkv,Bechtle:2020uwn,Bahl:2022igd}. The expected signal is computed by rescaling the corresponding production cross-sections and partial decay widths of the SM Higgs boson with the $\kappa_i$ parameters defined in the previous section. For masses above $m_\chi=1$~TeV we use a custom leading-order evaluation of the partial dilaton decay widths.
The estimated future $3 \,\text{ab}^{-1}$ HL-LHC sensitivity is obtained from the 13 TeV LHC analyses by rescaling the sensitivities with a square root of the corresponding luminosity ratios. Some of the currently most sensitive experimental analyses include~\cite{ATLAS:2020tlo,ATLAS:2018sbw,CMS:2021klu,ATLAS:2020fry,CMS:2018amk}, searching for heavy resonances produced in gluon fusion or vector boson fusion, and decaying into pairs of on- or off-shell EW vector bosons.

Let us now discuss the sensitivity of the dilaton collider phenomenology to the parameters discussed in the previous sections, and the resulting experimental bounds on them.  

\begin{figure}[t]
\centering
\includegraphics[width=7.8cm]{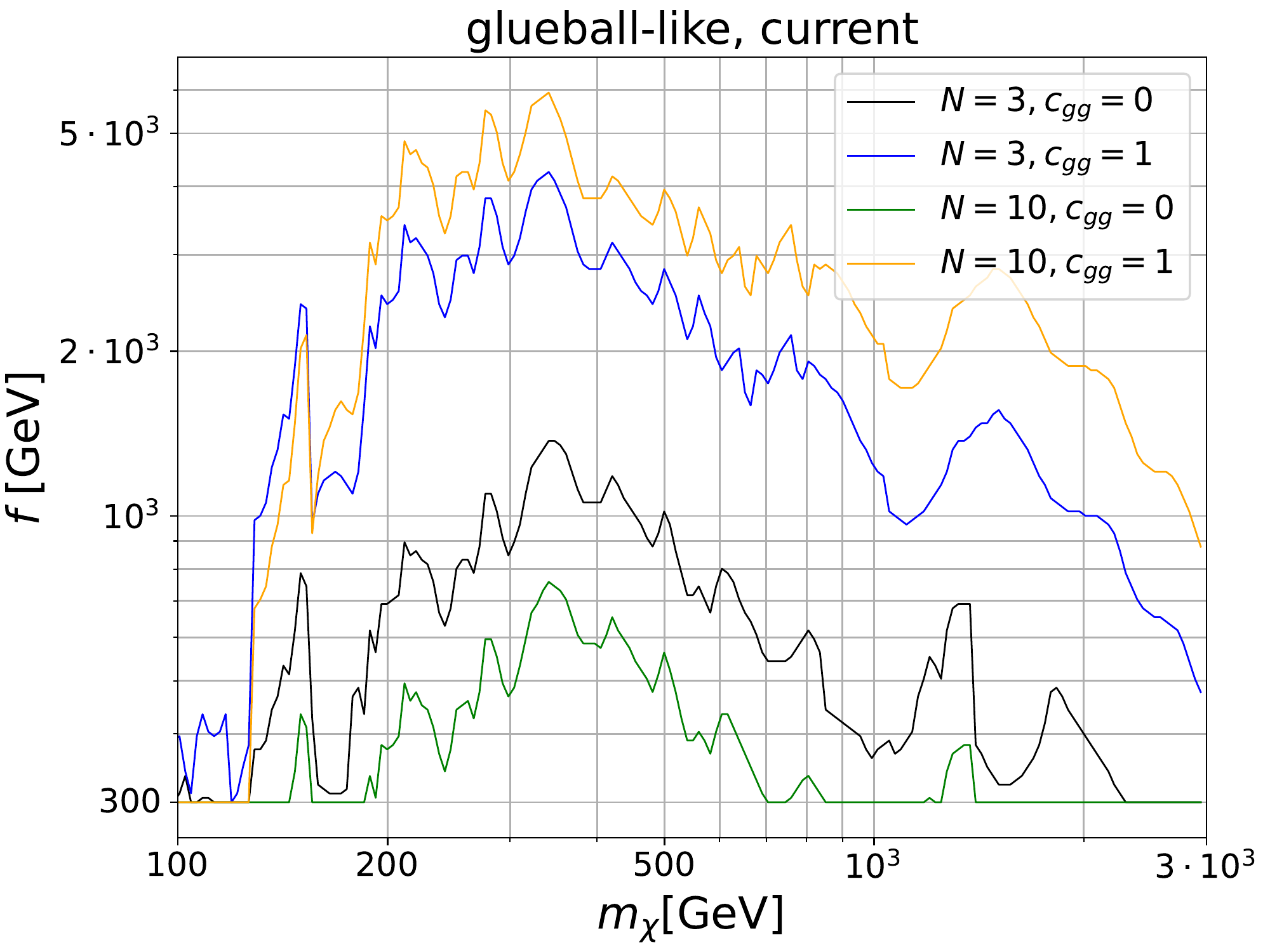}
\hspace{-0.3cm}
\includegraphics[width=7.8cm]{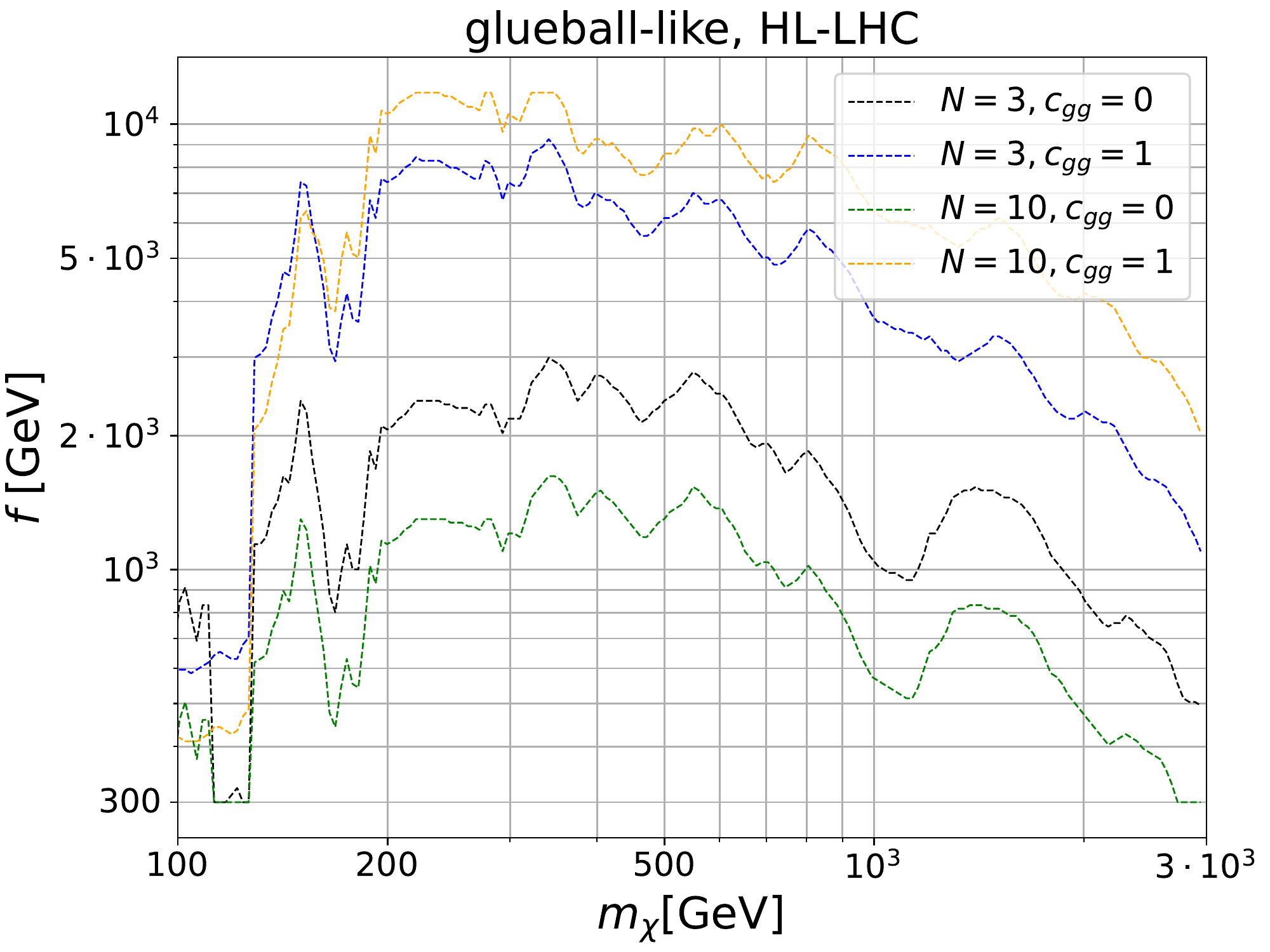} \\
\vspace{0.3cm}
\includegraphics[width=7.8cm]{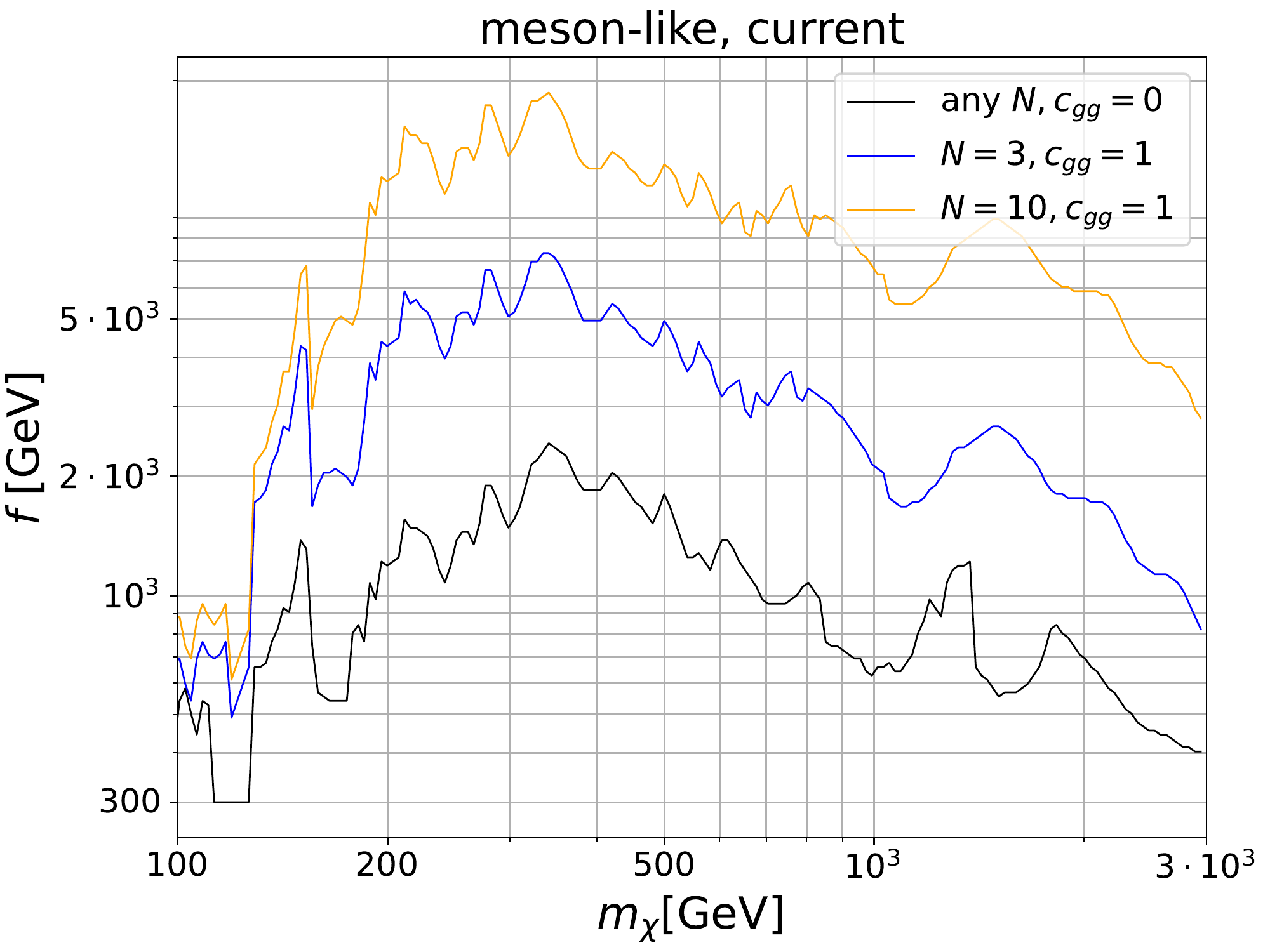}
\hspace{-0.3cm}
\includegraphics[width=7.8cm]{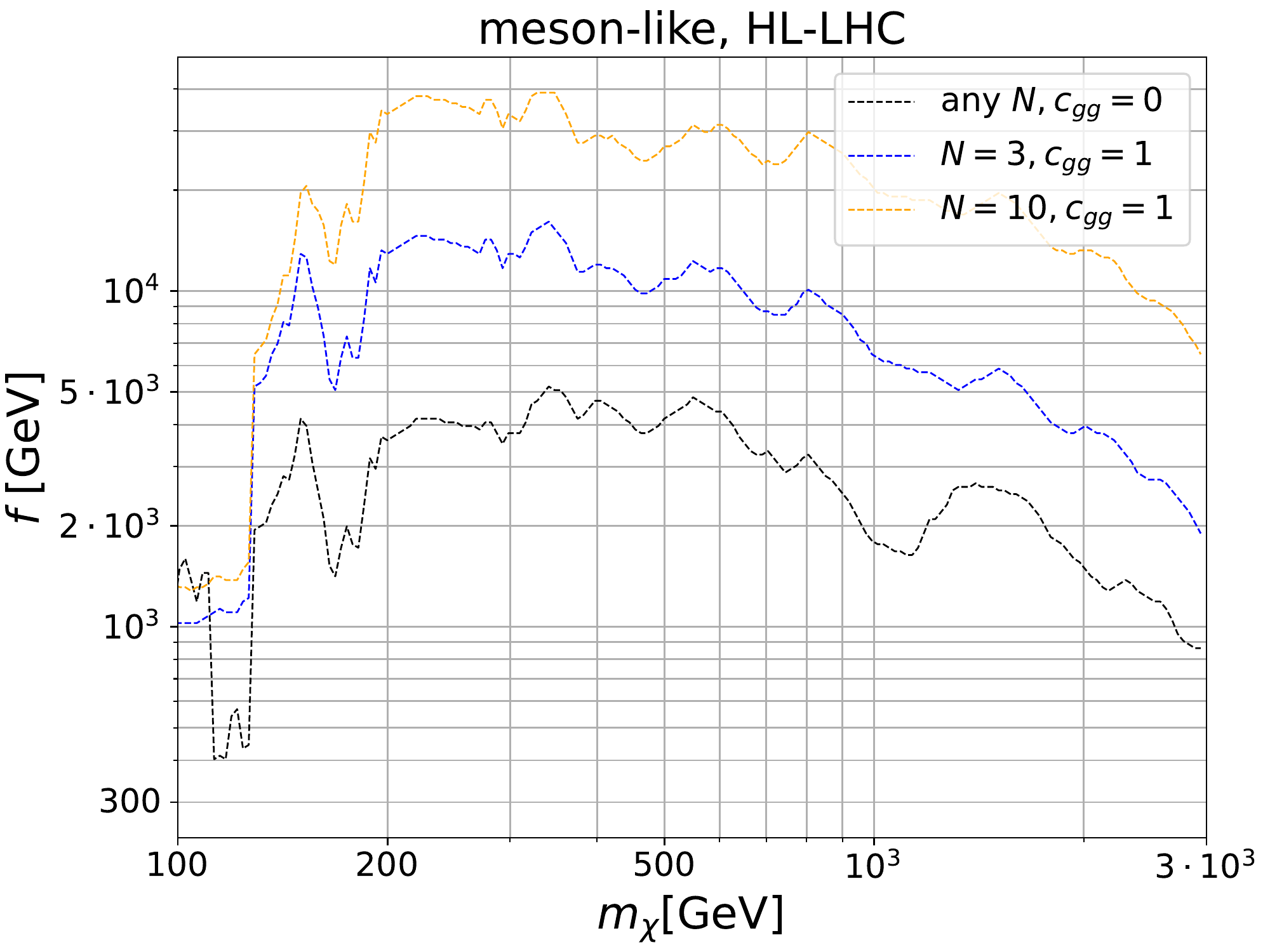}
\caption{\small \it{Current collider lower bounds on $f$ (left panel) and future sensitivity (right panel) as a function of the dilaton mass $m_\chi$ for a glueball- and meson-like dilaton, and for different choices of $c_{gg}$ and $N$ as specified in the plots. The other parameters are chosen as $c_{h \chi}=1$, $s_\theta=0$, $\gamma_i=0$, $c_{WW}= c_{BB}=0$.}}
\label{fig:plot1}
\end{figure}

\begin{itemize}

\item
The overall size of the dilaton couplings is set by the scale  of conformal symmetry breaking $\chi_0$, and the Higgs decay constant $f$, which are related via Eq.~(\ref{eq:twoscales}). While the experimental data provides lower bounds on $f$, EW scale naturalness pushes $f$ downwards. Hence expressing the bounds in terms of $f$ allows to estimate the degree of naturalness of the surviving region in parameter space. The bounds on $f$ and $m_\chi$ are shown in Fig.~\ref{fig:plot1}, for different choices of other relevant parameters which are discussed below. The ratio of $f$ and $\chi_0$ depends on the order-one parameter $c_{h\chi}$. To estimate the associated uncertainty we show in Fig.~\ref{fig:plot2} how the experimental bounds change as  $c_{h\chi}$ varies in the range $[1/2,2]$. The scaling of the bounds with $c_{h\chi}$ is discussed around Eqs.~(\ref{eq:bound1}) and (\ref{eq:bound2}) below.

\begin{figure}[t]
\centering
\includegraphics[width=7.7cm]{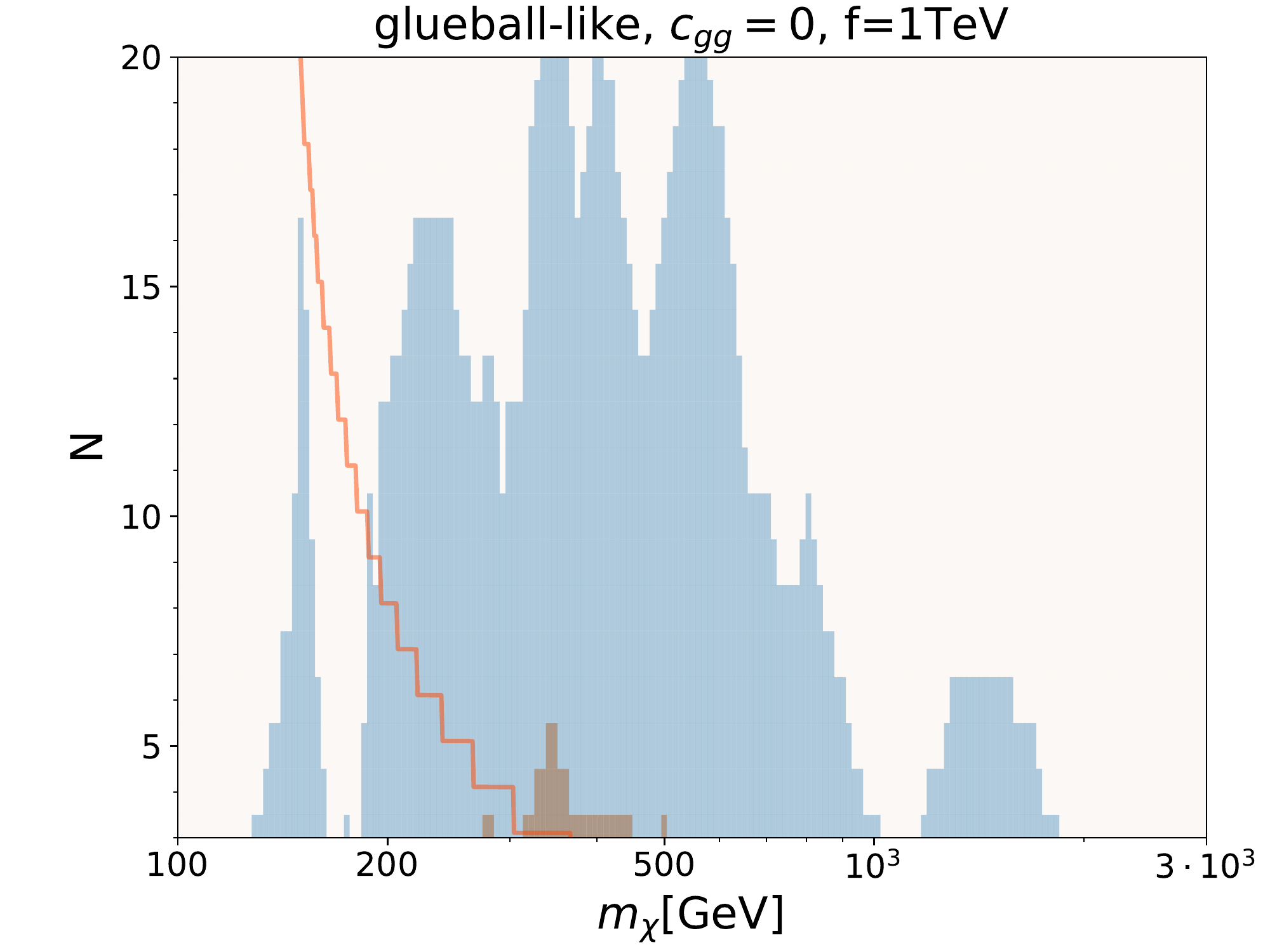}
\hspace{-0.3cm}
\includegraphics[width=7.7cm]{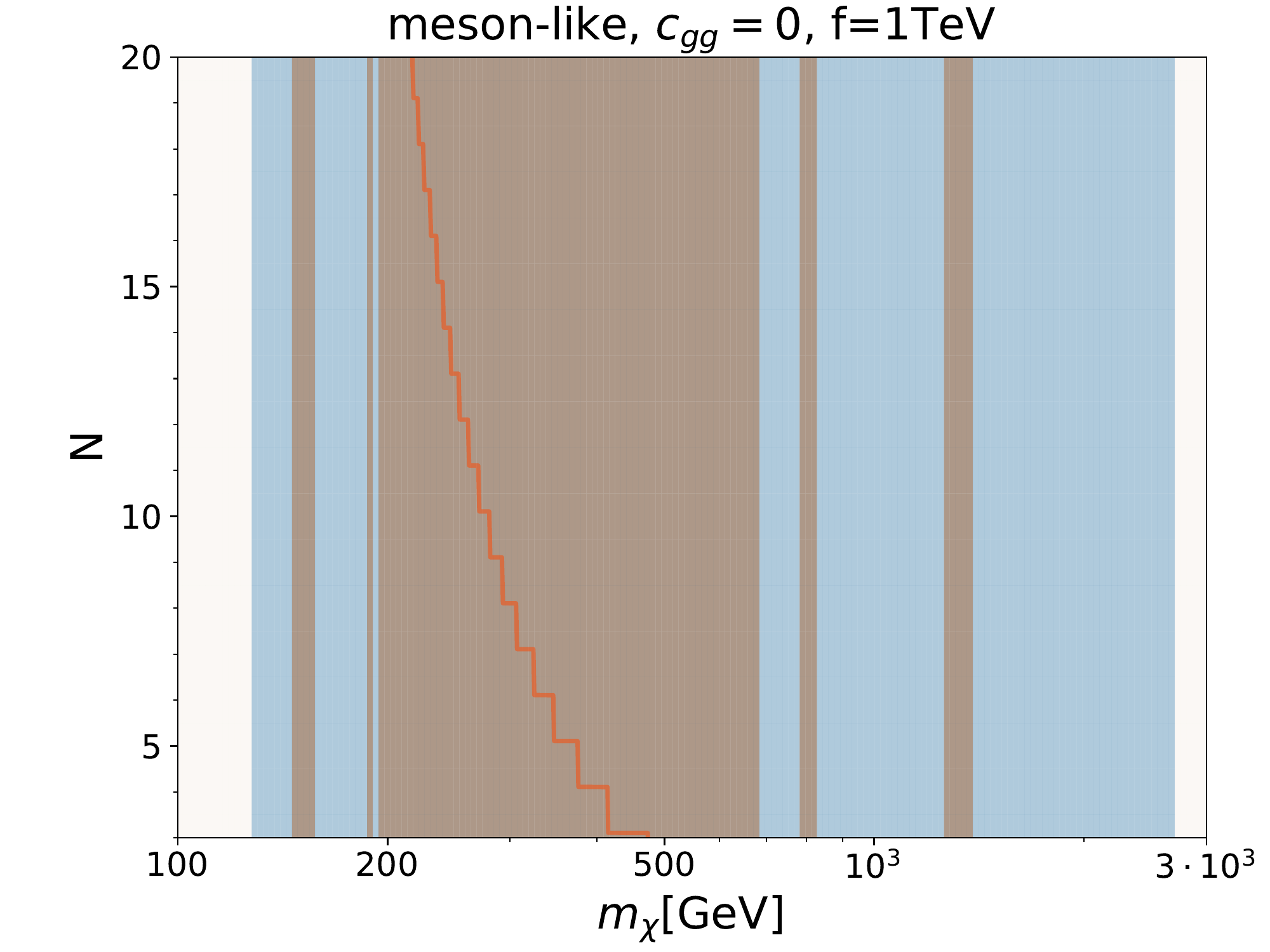} \\
\vspace{0.3cm}
\includegraphics[width=7.7cm]{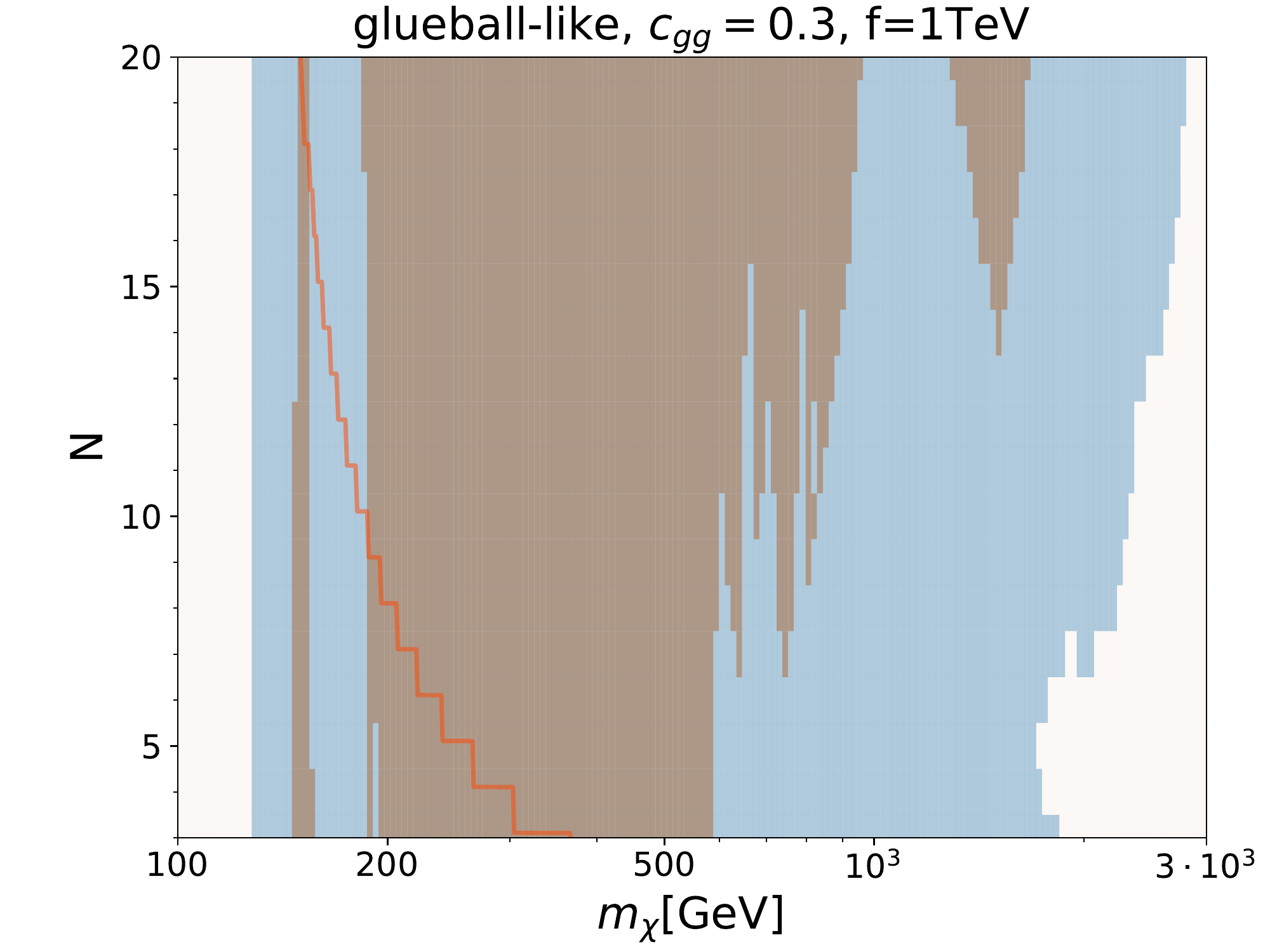}
\hspace{-0.3cm}
\includegraphics[width=7.7cm]{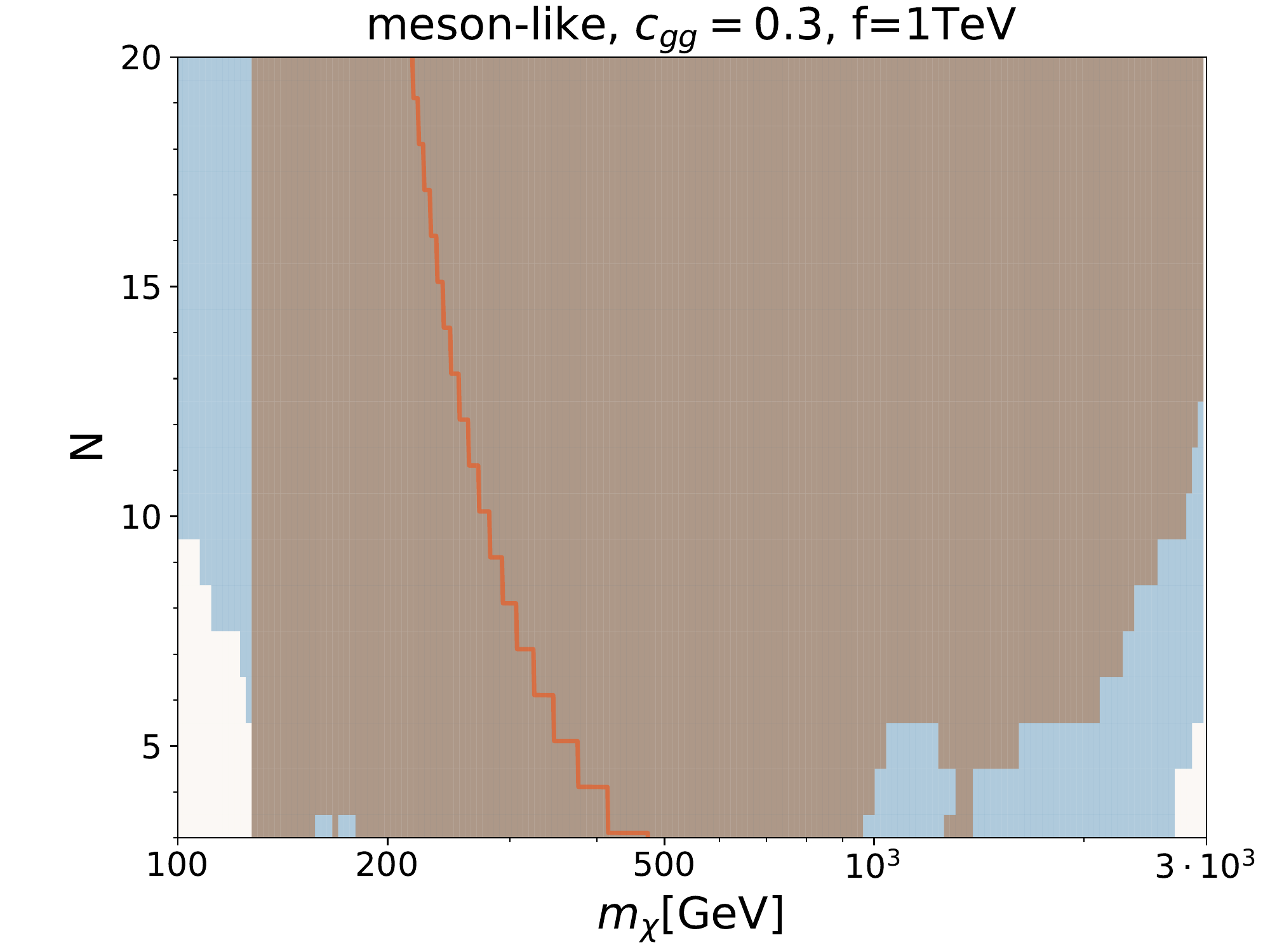}\\
\vspace{0.3cm}
\includegraphics[width=7.7cm]{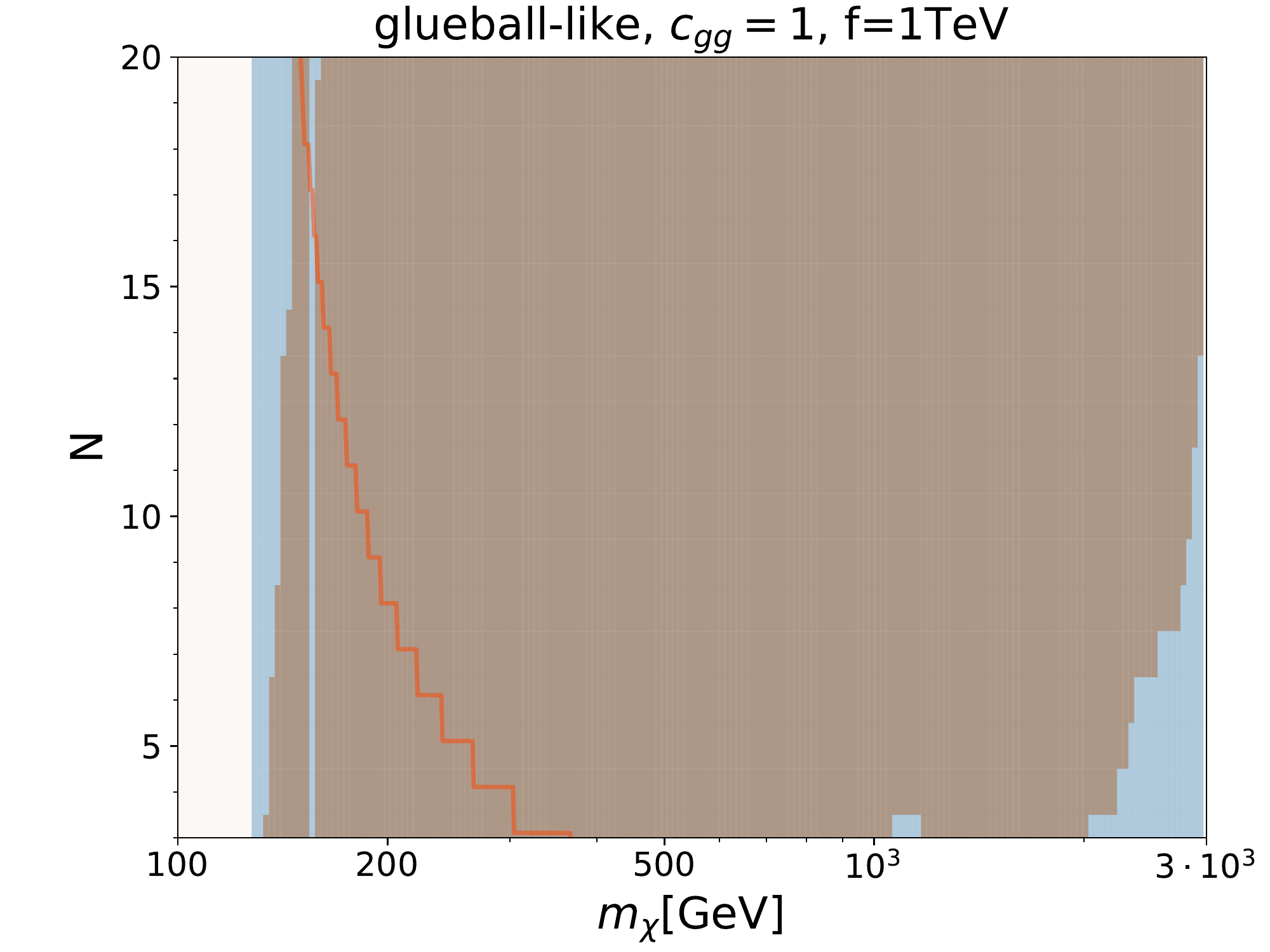}
\hspace{-0.3cm}
\includegraphics[width=7.7cm]{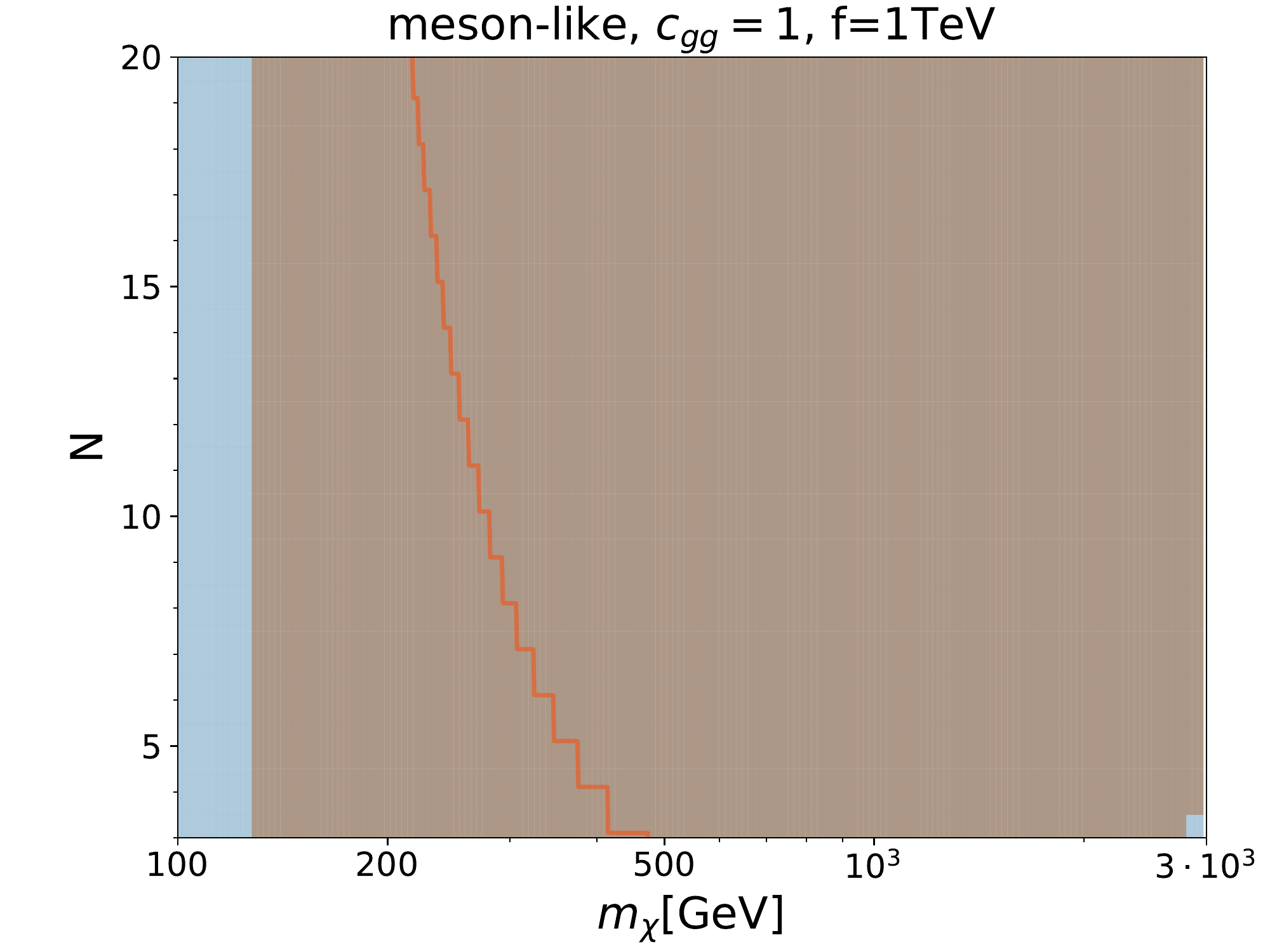}
\caption{\small \it{Currently excluded regions (brown) and future sensitivities (blue) in terms of the dilaton mass $m_\chi$ and $N$, for $c_{gg}=\{0,0.3,1\}$, $f=1$~TeV, $c_{h\chi}=1$, $s_\theta=0$, $\gamma_i=0$, and $c_{WW}= c_{BB}=0$. Red lines indicate the right edge of the region around $m_\chi = m_h$ excluded by the minimal mass splitting condition~(\ref{eq:minsplit}), with $\gamma_{\text{elem}}=0.1$. For $\gamma_{\text{elem}}=0$ the mass splitting condition only cuts out a thin region around $m_\chi=m_h$.}}
\label{fig:plot4}
\end{figure}

\item
The effective number of colors $N$ of the underlying new strong dynamics can {\it a priori} vary in a large range and plays a crucial role for the collider phenomenology. In particular, $N$ suppresses the overall scale of the couplings for a glueball-like dilaton $\propto 1/\chi_0 \propto 1/\sqrt N$ and, at the same time, enhances the dilaton coupling to gluons $\propto c_{gg} N$. The latter coupling determines the dominant dilaton production channel via gluon fusion (while the main decay channels are $\chi \to WW,ZZ$). The dependence of the bounds on $c_{gg}$ and $N$ is demonstrated in Fig.~\ref{fig:plot1}. As one can see, for non-zero $c_{gg}$ the experimental sensitivity to the dilaton grows significantly with $N$. In fact, the production cross-section from gluon fusion scales approximately as
\bea
\label{eq:bound1}
\sigma_{gg} \propto (c_{gg} N/\chi_0)^2 \propto \frac{1}{f^2} \frac{c_{gg}^2}{c_{h \chi}^2}\times
 \begin{cases}
 N &\; \mbox{for glueball-like dilaton}\\
N^2 & \; \mbox{for meson-like dilaton.}\\
\end{cases}
\eea
For vanishing $c_{gg}$ instead, dilaton production from gluon or vector boson fusion is mostly determined by the coupling to the top quark and the EW gauge bosons respectively. At zero mixing these couplings scale as $\propto 1/\chi_0$ and hence the overall production cross-section has the following $N$-dependence:
\bea
\label{eq:bound2}
\sigma_{gg,VBF}|_{c_{gg}=0} \propto (1/\chi_0)^2 \propto 
 \frac{1}{f^2} \frac{1}{c_{h \chi}^2}
 \times \begin{cases}
 \frac{1}{N} & \;\mbox{for glueball-like dilaton}\\
1 & \; \mbox{for meson-like dilaton.}\\
\end{cases}
\eea 
A meson-like dilaton is therefore typically more constrained than one that is glueball-like.
This can be seen in Fig.~\ref{fig:plot4}, where we show the experimental bounds in terms of $N$ and $m_\chi$ for a fixed value of $f$.

\item
The Higgs-dilaton mixing $s_\theta$ arises from operators which contain both Higgs-shift-symmetry and conformal-symmetry breaking parameters. The sources of this breaking are model-dependent. Since the current Higgs-coupling bounds prefer a SM-like Higgs~\cite{CMS:2022dwd,ATLAS:2022vkf}, we have set the mixing to zero for most of the plots, which allows to satisfy the bounds for all $f\gtrsim 800$~GeV (see Fig.~\ref{fig:plothVV}). 
To estimate the importance of the mixing for the dilaton collider bounds we present in Fig.~\ref{fig:plot3} a comparison of the bounds derived for $s_{\theta}=0,\pm 0.05$.

\begin{figure}[t]
\centering
\includegraphics[width=7.8cm]{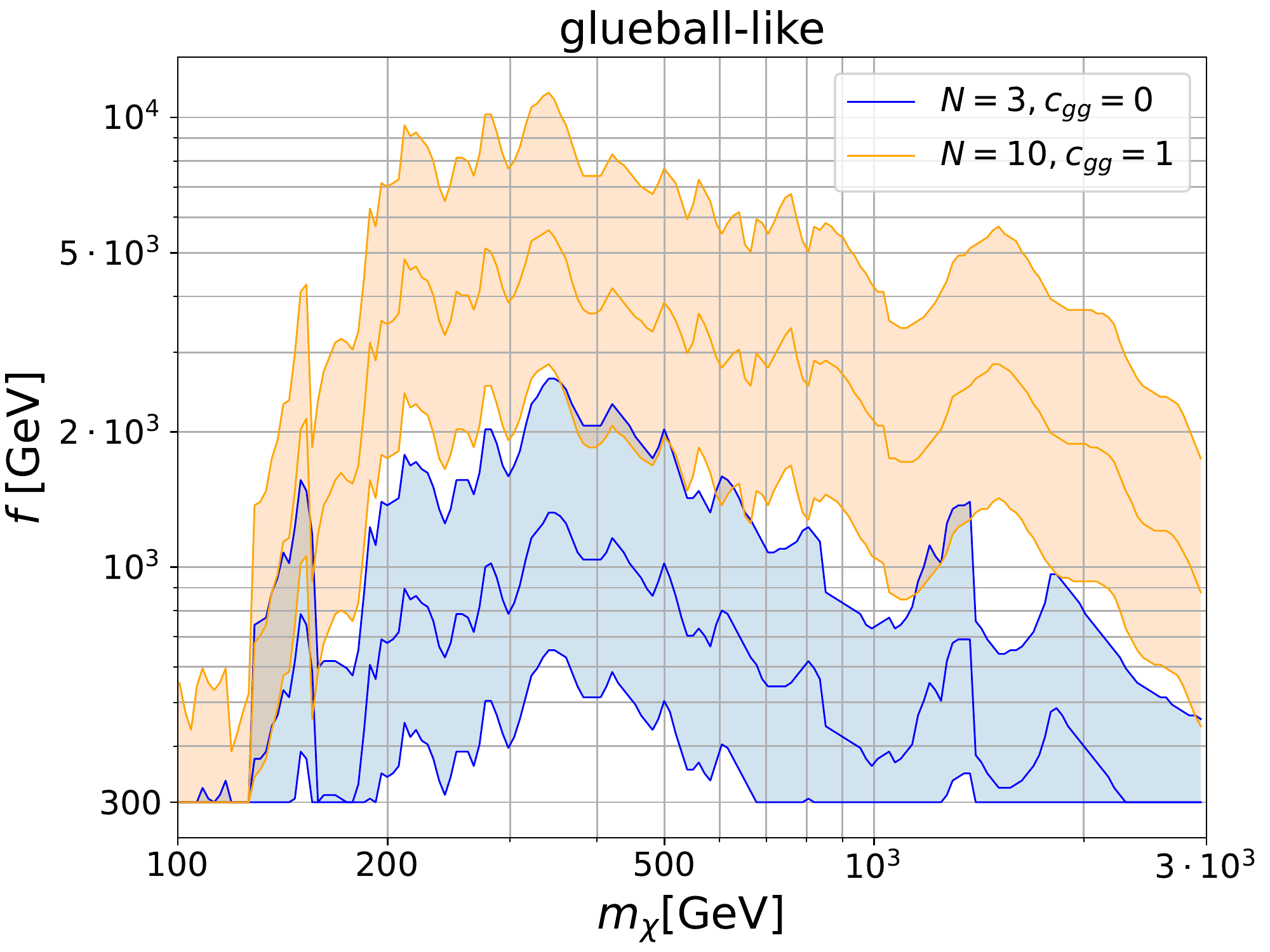}
\hspace{-0.3cm}
\includegraphics[width=7.8cm]{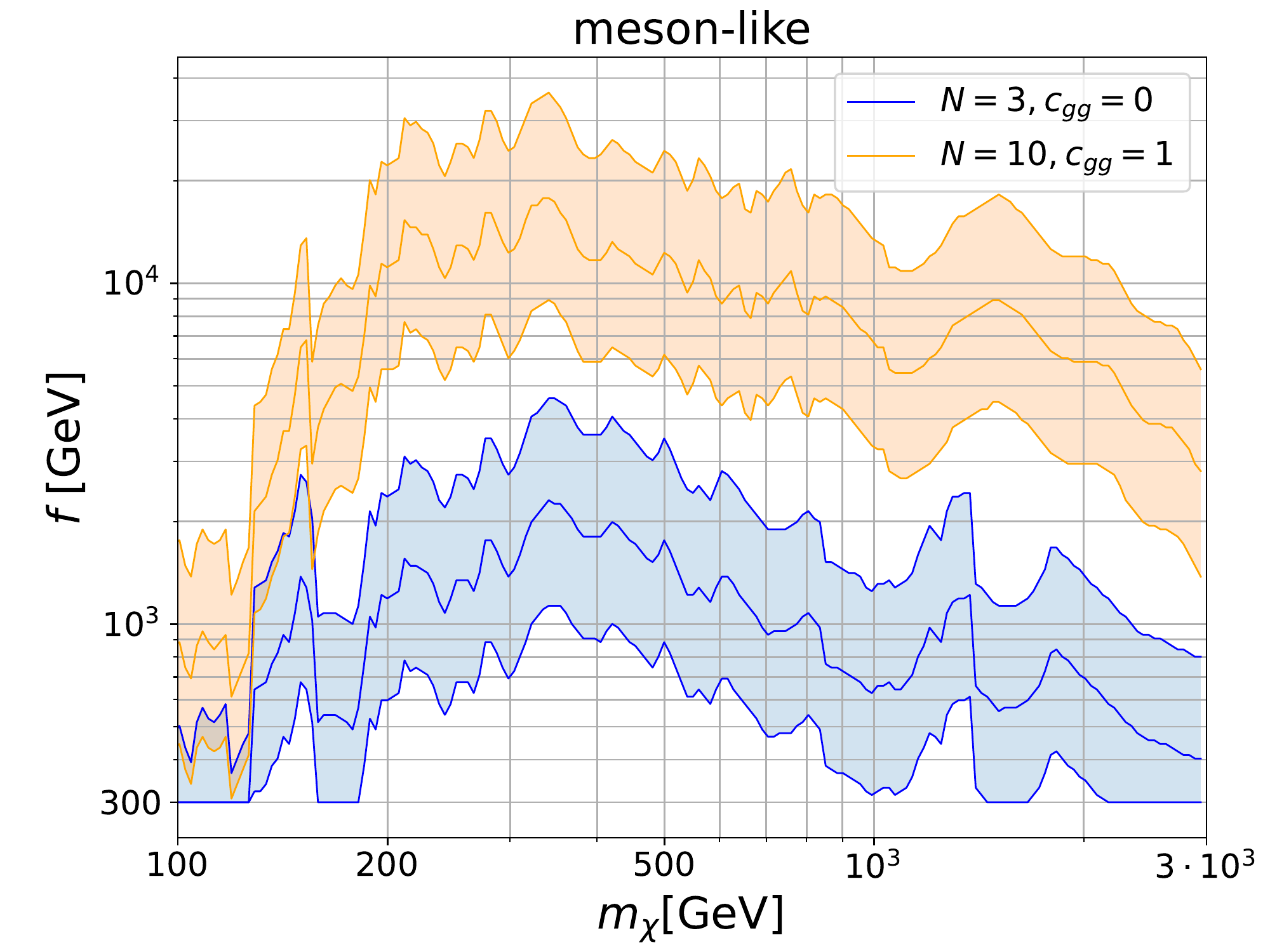}
\caption{\small \it{Dependence of the currently excluded $f$ on $c_{h\chi}$ varying between $1/2$ and $2$. The upper (lower) band limits correspond to $c_{h\chi}=1/2\,(2)$, the central lines correspond to $c_{h\chi}=1$.  The values of $c_{gg}$ and $N$ are specified in the plots, the other parameters are chosen as $s_\theta=0$, $\gamma_i=0$, $c_{WW}= c_{BB}=0$. 
}}
\label{fig:plot2}
\end{figure}
\begin{figure}[t]
\centering
\includegraphics[width=7.8cm]{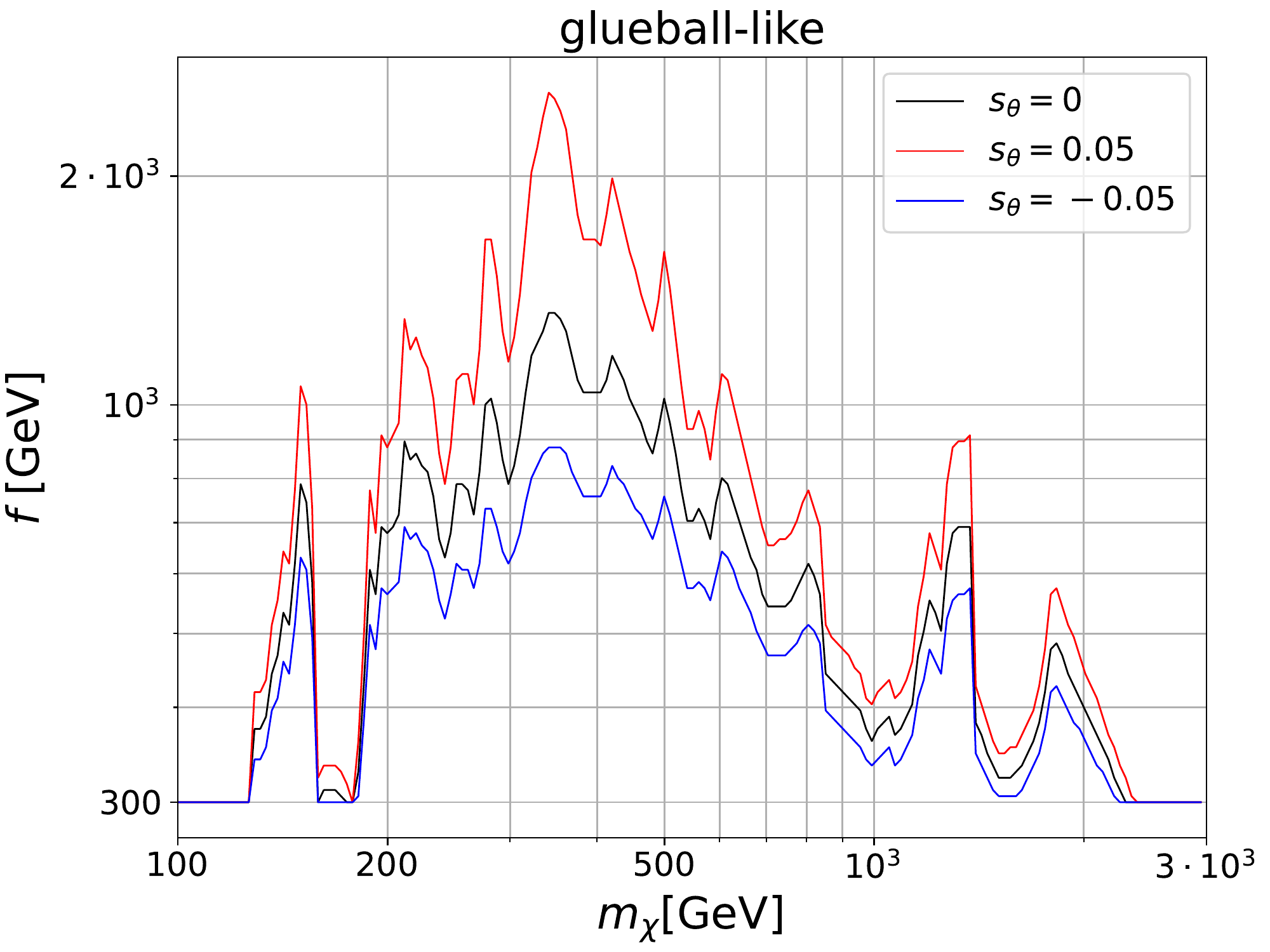}
\hspace{-0.3cm}
\includegraphics[width=7.8cm]{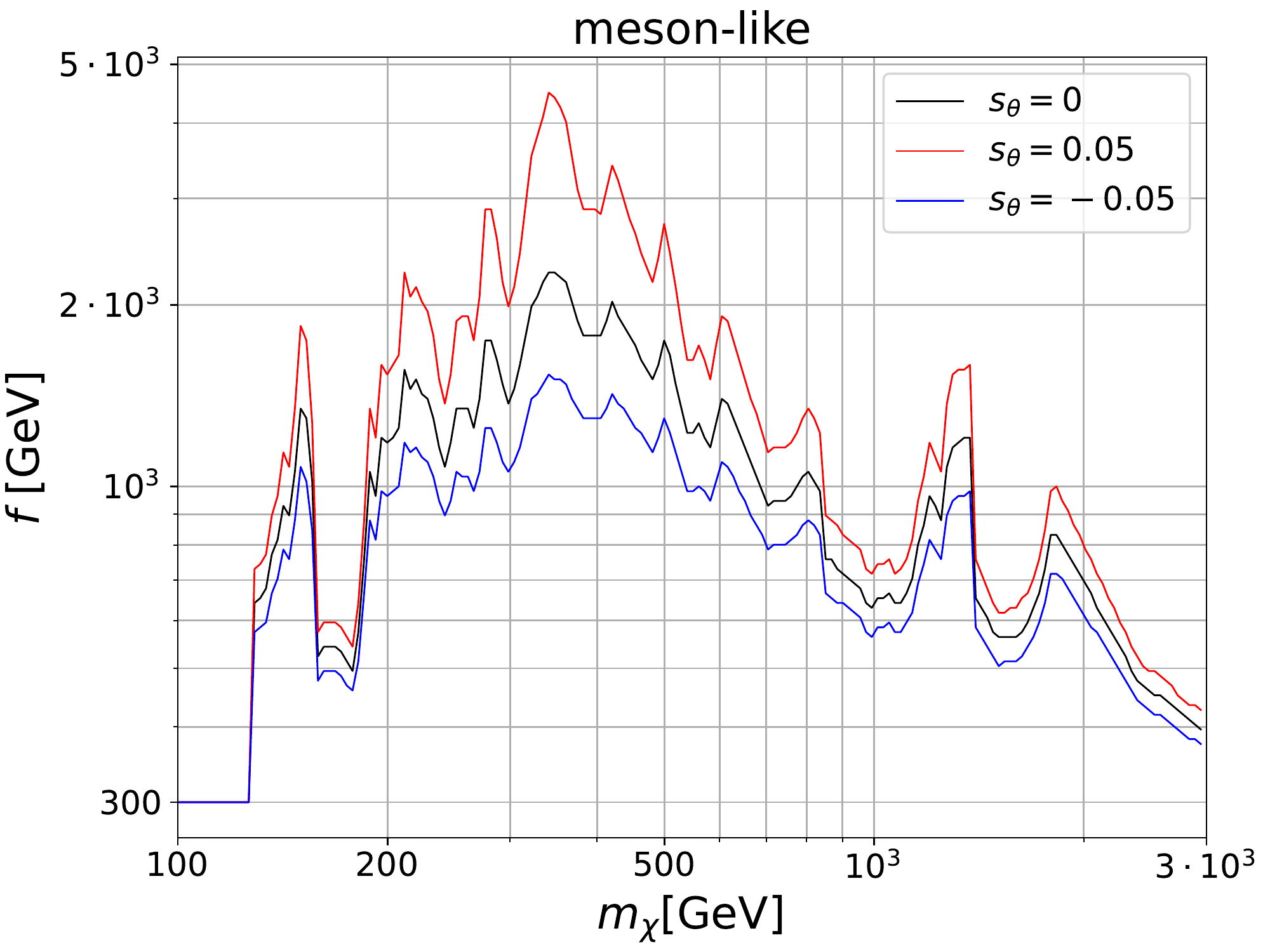}
\caption{\small \it{Dependence of the currently excluded $f$ on the Higgs-dilaton mixing angle $s_\theta$, for $s_\theta = 0, \pm 0.05$. The other parameters are chosen as $c_{h\chi}=1$, $c_{gg}=0$, $N=3$, $\gamma_i=0$, $c_{WW}= c_{BB}=0$. }}
\label{fig:plot3}
\end{figure}

\begin{figure}[t]
\centering
\includegraphics[width=7.8cm]{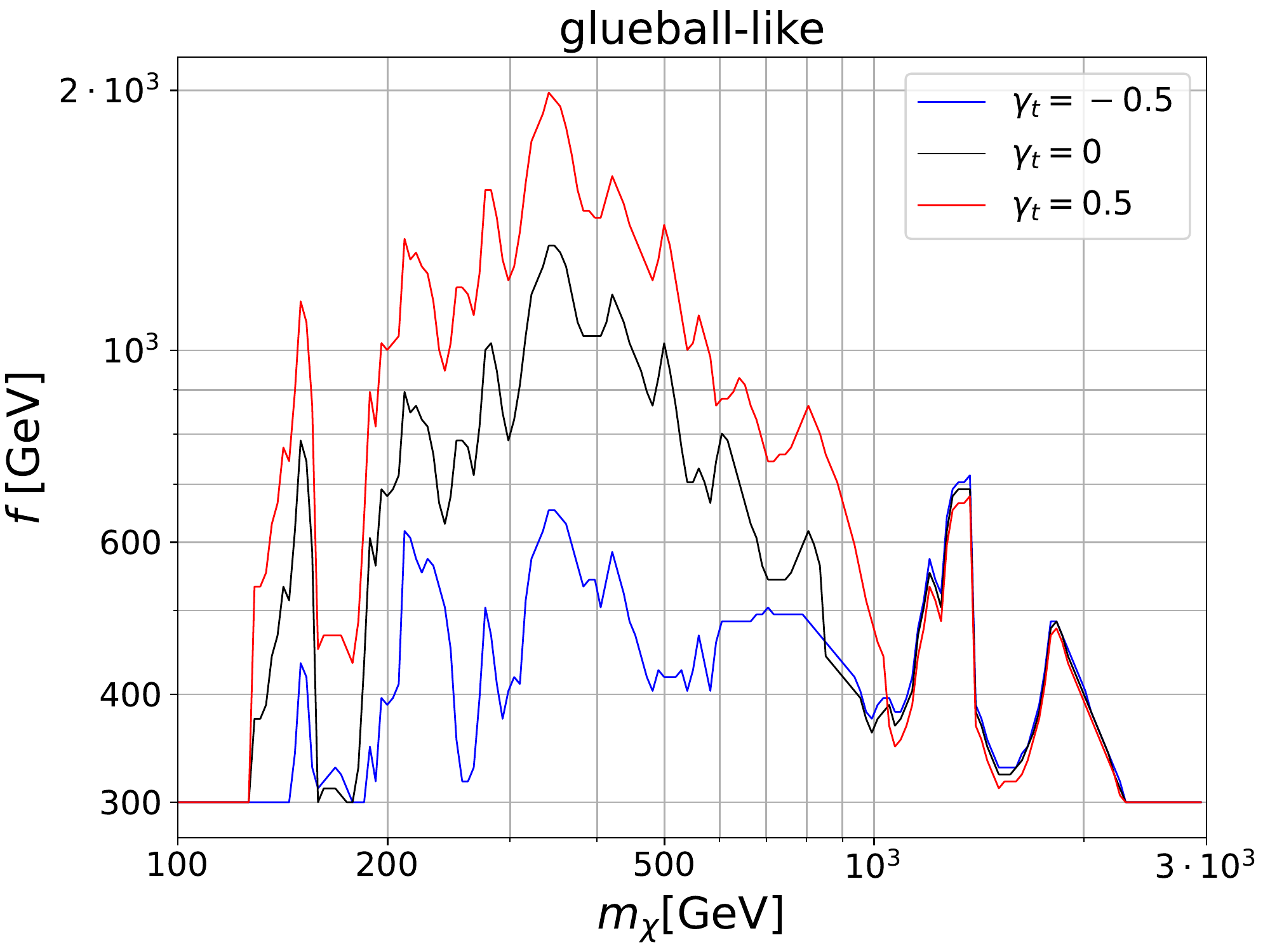}
\hspace{-0.3cm}
\includegraphics[width=7.8cm]{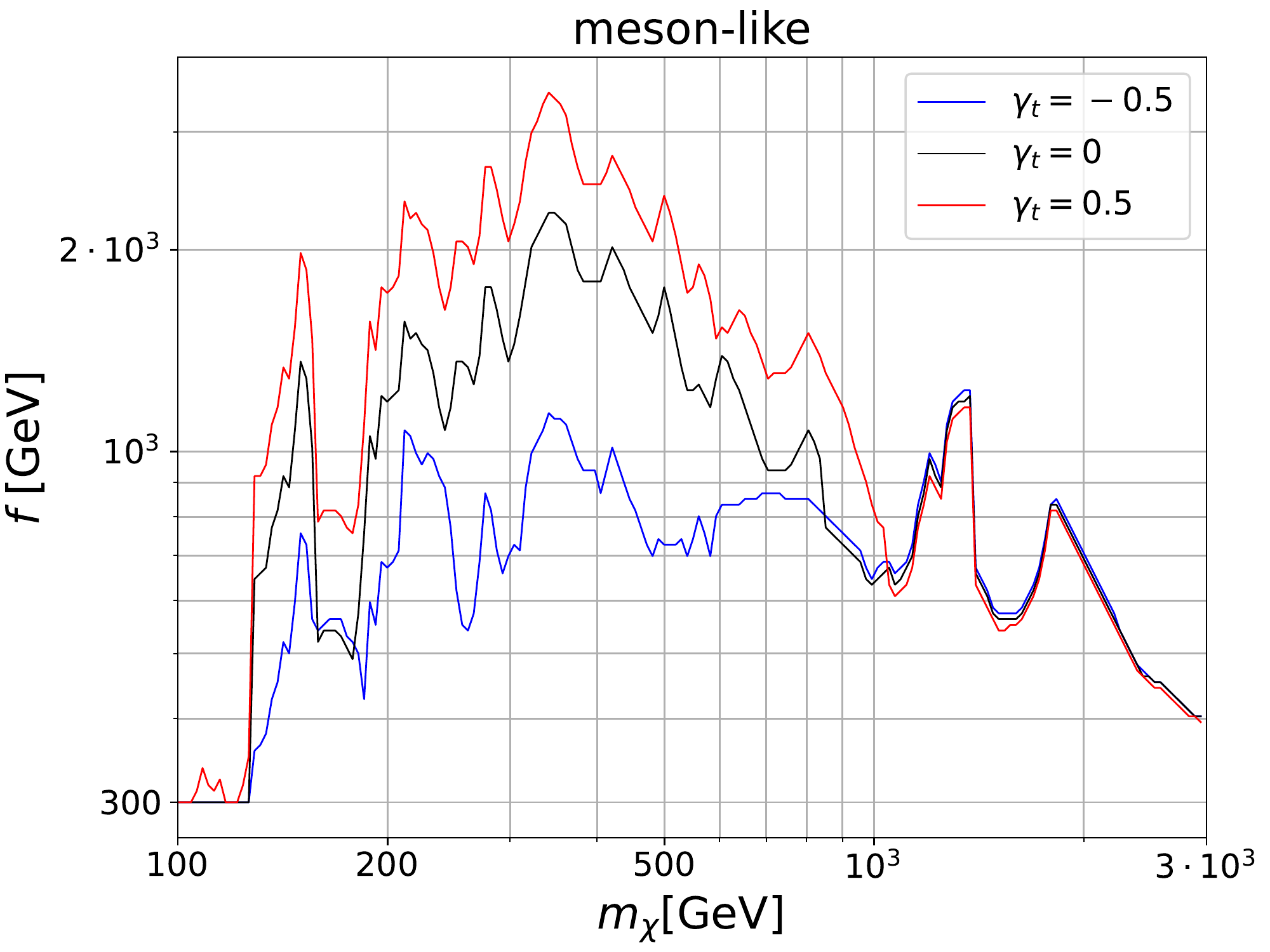}
\caption{\small \it{Dependence of the currently excluded $f$ on $\gamma_t$, for $\gamma_t = 0, \pm 0.5$, for a glueball-like (left panel) and a meson-like (right panel) dilaton. The remaining parameter values are chosen as $c_{h\chi}=1$, $c_{gg}=0$, $N=3$, $s_\theta=0$, $\gamma_i=0$, $c_{WW}= c_{BB}=0$. 
}}
\label{fig:plot5}
\end{figure}

\item
Regarding the order-one parameters $c_{gg}$, $c_{WW}$, $c_{BB}$, the latter two have a very mild impact on the collider sensitivity and we will set them to zero. The former, instead, can play an important role due to its effect on the coupling to gluons, as was discussed above. The dependence on $c_{gg}$ is demonstrated in Figs.~\ref{fig:plot1},~\ref{fig:plot4},~\ref{fig:plot2}.

\item
The parameters $\gamma_\psi, \gamma_{V^2}$ which reflect the scale-invariance breaking in various couplings were set to zero in our analysis as they are model-dependent. Among them, the most important one is the parameter $\gamma_{t}$, whose order-one value could affect the dilaton-top coupling, and hence the dilaton coupling to gluons, modifying the overall production rate. This effect is more sizeable for low $c_{gg} N$, since otherwise the latter contribution dominates the coupling to gluons. This is illustrated in Fig.~\ref{fig:plot5}.

\end{itemize}

\section{Discussion}\label{sec:disc}

The existence of a ${\cal O}$(100 GeV) -- ${\cal O}$(1 TeV) scale 
dilaton was predicted in numerous  extensions of the SM, and in particular in models with Higgs boson compositeness.  
In this work, we have presented an update of the collider bounds on a composite dilaton with mass $\gtrsim 100$~GeV, as well as the projected sensitivity of the HL-LHC. We have assumed the dilaton to be the lightest new state, decaying exclusively to SM particles. To model the dilaton properties we have used a 4D EFT approach, being able to capture a broad range of possible UV-completions, while still allowing to put important restrictions on the dilaton interactions.

We have adopted a perspective on the dilaton phenomenology motivated by EW scale naturalness, relating the dilaton couplings to the Higgs decay constant $f$ (which for the least tuned models currently lies within about $1$\,TeV), and the number of colors $N$ of the underlying new strong interactions producing the composite Higgs and dilaton. 
Important relations to understand the constraints are Eqs.~(\ref{eq:twoscales}), (\ref{eq:bound1}) and (\ref{eq:bound2}).
In particular, the non-trivial relation between the meson and glueball VEVs implies that a large scale $\chi_0$ suppressing the dilaton couplings can only be achieved at the price of large $N$. At the same time, this large $N$ is expected to enhance the dilaton couplings to gluons, increasing the resulting signal. As a result, at fixed $f\lesssim 1$~TeV, one expects sizeable collider signals for a glueball-like dilaton even at large $\chi_0$.

The above line of arguments is only valid for non-zero values of the coefficient $c_{gg}$ controlling the CFT contribution to the QCD $\beta$-function and thus the dilaton coupling to gluons. However, having a non-vanishing $c_{gg}$ appears to be a rather generic assumption, since the CFT constituents have to be charged under QCD in order to generate at least the top quark mass via the partial compositeness mechanism, which requires QCD-charged composite operators coupling to the top quark.

Finally, in the hypothetical case of a meson-like dilaton, the scale controlling the dilaton interactions is rigidly connected to $f$ and does not grow with $N$. Hence, the experimental sensitivity in this case is much stronger than in the glueball scenario.

This motivates LHC searches for a heavy dilaton. The main production channel benefiting from the mentioned parametric dependence is $gg \to \chi$, followed by decays to the EW gauge bosons.
As we have shown, a dilaton with mass up to a few TeV can be observed or constrained at future LHC runs. On the other hand, the current experimental data already excludes a significant fraction of the parameter space for a composite dilaton, advancing well into the TeV mass region. In particular, the obtained bounds are of special relevance for models of electroweak baryogenesis~\cite{Bruggisser:2018mus,Bruggisser:2018mrt,us:ewbg}, even in realisations where the temperature of the electroweak phase transition is enhanced~\cite{Baldes:2018nel,Matsedonskyi:2020mlz,Matsedonskyi:2020kuy,Matsedonskyi:2021hti,us:snr}, which rely on the presence of a \smash{${\cal O}$(100 GeV)} -- ${\cal O}$(1 TeV) scale dilaton.
At the same time, the window of a light dilaton with a mass $\lesssim 200$~GeV remains only weakly constrained, with the latest ATLAS and CMS analyses concentrating on heavier masses. Although a fraction of this window can be excluded if large Higgs-dilaton mixing is generated (see Fig.~\ref{fig:plot4}), in general it represents a theoretically-acceptable region of parameter space and would require a dedicated effort to be probed experimentally. 

Finally, the analysed connection between the Higgs and the dilaton scales provides a handle to assess the implications of dilaton searches on EW scale naturalness. In the considered benchmark scenarios the current exclusion limits and the future sensitivity reach respectively 5 TeV and 10 TeV in the Higgs decay constant $f$, corresponding to a $\sim 10^{-4}$ level of fine-tuning. This sensitivity, although being model-dependent, can exceed the sensitivity provided by the measurements of the couplings of SM particles at the HL-LHC and next generation lepton colliders~\cite{Durieux:2018ekg,Cepeda:2019klc}.
Although, simplistically comparing the dilaton production cross section which scales as $\propto 1/f^2$ with the SM-coupling deviations induced by compositeness $\propto 1/f^2$, we find them to be of the same parametric form, the dilaton signal profits from the $N$-enhancement of the gluon fusion cross-section.  
Thus dilaton searches can complement the program of Higgs compositeness tests, including such directions as the searches for composite top partners~\cite{Contino:2008hi,Matsedonskyi:2014lla,Matsedonskyi:2015dns}, precision tests of SM particle interactions~\cite{Grojean:2013qca,Matsedonskyi:2014iha,Cacciapaglia:2022tfd}, and searches for other types of light spin-zero composite states~\cite{Franceschini:2015kwy,Chala:2017sjk,Cacciapaglia:2019bqz,Cornell:2020usb}. In case a new light boson is discovered, the dilaton coupling patterns analysed in this paper as well as the patterns predicted for other types of composite resonances~\cite{Franceschini:2015kwy,Chala:2017sjk} would provide a useful guide for understanding the origin of such a state.

\section*{Acknowledgments}

We thank Miki Chala for early discussions on this topic a few years ago.
OM is supported by STFC HEP Theory Consolidated grant ST/T000694/1. OM also thanks the Mainz Institute for Theoretical Physics (MITP) and ICTP-SAIFR for their hospitality and support during the completion of this work. 
 The work of SB  has been supported by the German Research Foundation (DFG) under grant no.~396021762–TRR 257.~This work is supported by the Deutsche Forschungsgemeinschaft under Germany Excellence Strategy - EXC 2121 ``Quantum Universe'' - 390833306.

\appendix
\section{Loop Functions} \label{sec:loopfunctions}
	
In this appendix we list the loop functions presented in Ref.~\cite{Spira:2016ztx}:
\bea
A_q(\tau) &=& \frac 3 2 \tau (1+(1-\tau)f(\tau)) \\
A_f(\tau) &=&  2 \tau (1+(1-\tau)f(\tau)) \\
A_W(\tau) &=& -(2+3 \tau + 3 \tau (2-\tau)f(\tau)) 
\eea
\bea
A_f(\tau,\lambda) &=& 2 N_{cf} \frac{e_f(I_{3f}-2 e_f s_w^2)}{c_w} (I_1(\tau,\lambda)-I_2(\tau,\lambda)) \\
A_W(\tau,\lambda) &=& c_w ( 4(3-t_w^2)I_2(\tau,\lambda) + ((1 + 2/\tau)t_w^2 - (5+2/\tau)) I_1(\tau,\lambda))
\eea
\bea
I_1(\tau,\lambda) &=& \frac{\tau \lambda}{2(\tau-\lambda)} + \frac{\tau^2\lambda^2}{2(\tau-\lambda)^2}(f(\tau)-f(\lambda)) + \frac{\tau^2 \lambda}{(\tau-\lambda)^2}(g(\tau)-g(\lambda))\\
I_2(\tau,\lambda) &=& - \frac{\tau \lambda}{2(\tau-\lambda)}(f(\tau)-f(\lambda)).
\eea
Here $t_w=s_w/c_w$, $e_f$ is the electric charge and $I_{3f}$ the third electroweak isospin component of the corresponding fermion. 
Furthermore,
\bea
f(\tau) &=&     \begin{cases}
      \arcsin^2 \frac 1 {\sqrt \tau} & \tau\geq 1\\
      - \frac 1 4 (\log \frac{1+ \sqrt{1-\tau}}{1-\sqrt{1-\tau}}-i \pi)^2 & \tau < 1
    \end{cases} \\
    g(\tau) &=&     \begin{cases}
      \sqrt{\tau-1}\arcsin \frac 1 {\sqrt \tau} & \tau\geq 1\\
      \frac{\sqrt{1-\tau}}{2} (\log \frac{1+ \sqrt{1-\tau}}{1-\sqrt{1-\tau}}-i \pi) & \tau < 1.
    \end{cases}
\eea

\newpage

\bibliographystyle{JHEP}  
\bibliography{biblio}

\end{document}